\documentclass[runningheads]{llncs}
\usepackage[T1]{fontenc}
\usepackage{graphicx}
\usepackage{hyperref}
\usepackage{comment}
\usepackage{hyperref}
\usepackage{amssymb}
\usepackage{subcaption}
\usepackage{multirow}
\usepackage{enumerate}
\usepackage{graphics}
\usepackage{amsmath,algorithm,algpseudocode,amsfonts}
\usepackage{makecell}
\usepackage{multicol}
\usepackage{hhline}
\usepackage{setspace}
\usepackage[group-separator=]{siunitx}

\usepackage{mathtools}

\allowdisplaybreaks[1]

\usepackage{tikz}
\usepackage{tikzsymbols}
\usepackage{fancyvrb}
\definecolor{mygrey}{rgb}{0.8, 0.8, 0.8}
\usetikzlibrary{arrows.meta}
\usetikzlibrary{calc}
\usetikzlibrary{chains}
\usetikzlibrary{decorations.pathreplacing}
\usetikzlibrary{positioning}
\usetikzlibrary{shapes}
\tikzstyle{arrow} = [draw, -latex']
\tikzset{
    rect/.style={align=center, bottom color=white, top color=white, draw=black, font=\small, minimum size=6mm, minimum width=1mm, node distance=0mm, outer sep=0mm, rectangle},
    empty/.style={align=center, bottom color=white, top color=white, draw=white, font=\small, minimum size=1mm, minimum width=1mm, node distance=0mm, outer sep=0mm, rectangle}
}

\newcommand{\exor}{XOR\xspace}
\newcommand{\sbx}{SBox\xspace}
\newcommand{\sbxs}{SBoxes\xspace}
\newcommand{\mc}{\texttt{MixColumn}\xspace}

\begin{document}
\title{Modeling Linear and Non-linear Layers: An MILP Approach Towards Finding Differential and Impossible Differential Propagations}
\titlerunning{Finding Differential and Impossible Differential Propagations}
%
\author{Debranjan Pal \and
Vishal Pankaj Chandratreya \and
Abhijit Das\and\\
Dipanwita Roy Chowdhury}
\authorrunning{D Pal et al.}
%
\institute{Crypto Research Lab, \\Indian Institute of Technology Kharagpur, India\\
\email{debranjanpal@iitkgp.ac.in,}
\email{vpaijc@kgpian.iitkgp.ac.in}\\
\email{\{abhij,drc\}@cse.iitkgp.ac.in}}
\maketitle              
\begin{abstract}
Symmetric key cryptography stands as a fundamental cornerstone in ensuring security within contemporary electronic communication frameworks. The cryptanalysis of classical symmetric key ciphers involves traditional methods and techniques aimed at breaking or analyzing these cryptographic systems. In the evaluation of new ciphers, the resistance against linear and differential cryptanalysis is commonly a key design criterion. The wide trail design technique for block ciphers facilitates the demonstration of security against linear and differential cryptanalysis. Assessing the scheme's security against differential attacks often involves determining the minimum number of active SBoxes for all rounds of a cipher. The propagation characteristics of a cryptographic component, such as an SBox, can be expressed using Boolean functions. Mixed Integer Linear Programming (MILP) proves to be a valuable technique for solving Boolean functions. We formulate a set of inequalities to model a Boolean function, which is subsequently solved by an MILP solver. To efficiently model a Boolean function and select a minimal set of inequalities, two key challenges must be addressed. We propose algorithms to address the second challenge, aiming to find more optimized linear and non-linear components. Our approaches are applied to modeling SBoxes (up to six bits) and EXOR operations with any number of inputs. Additionally, we introduce an MILP-based automatic tool for exploring differential and impossible differential propagations within a cipher. The tool is successfully applied to five lightweight block ciphers: Lilliput, GIFT64, SKINNY64, Klein, and MIBS.
\end{abstract}

\section{Introduction}
Differential cryptanalysis~\cite{DifferentialCrypt_BihamShamirDCDES} and linear cryptanalysis~\cite{LinearCrypt_MatsuiY92} stand out as the primary techniques in the world of symmetric-key cryptography analysis. Differential cryptanalysis unveils the transformation of input differences in plaintext into output differences in ciphertext for block ciphers. In contrast, linear cryptanalysis expresses the probabilistic linear relationships among plaintext, ciphertext, and key. When scrutinizing a cipher's robustness, especially in the context of key-recovery attacks or distinguishers, the deviation of an ideal cipher from a random one becomes a pivotal feature for differential and linear cryptanalysis.

In the realm of new ciphers, it's common practice to prioritize resistance against linear and differential cryptanalysis during the cipher analysis process. The wide trail design technique~\cite{wide_trail_Daemen} for block ciphers is instrumental in proving security against these cryptanalysis methods. Evaluating the security of a scheme against differential attacks often involves determining the minimum number of active SBoxes within the cipher. While this task has garnered significant attention, the manual and computational effort required for its execution remains substantial. Enter MILP, a powerful optimization tool that deals with integral variables and offers an efficient approach to tackle complex problems.

To address the challenge of finding the minimum number of active SBoxes, attackers employ MILP to define potential differential propagation patterns in a round function. The subsequent execution of an MILP solver in parallel provides the minimal number of active SBoxes for the specified propagation patterns. Establishing a lower bound for the quantity of active SBoxes in differential and linear cryptanalysis allows for the calculation of an upper bound for the likelihood of the best characteristic. The maximum differential probability (MDP) of the SBoxes plays a crucial role in this process, enabling the accurate calculation of the overall differential probability by summing up the probabilities of all matching characteristics.

The propagation characteristic of cryptographic components, such as SBoxes, is often expressed through Boolean functions. Modeling these functions using the MILP method involves computing a set of inequalities and determining solutions that precisely correspond to supporting the Boolean function. Researchers grapple with the challenge of efficiently modeling Boolean functions by creating a minimal set of possible inequalities. Various techniques proposed by researchers, including those in references~\cite{Mouha,Sun_asiacrypt,sun_towards,Sun_SbP,BouraC,Sasaki_MILP_ALG,Superball_Li_Sun_tosc}, aim to address these challenges.

While MILP models are frequently employed in cryptanalysis, the quest for the most effective comprehensive model continues. Researchers emphasize the need to build and thoroughly investigate different proposed models, considering the variety of possible inequalities involved. The optimization of MILP models hinges on observing a reduction in the number of inequalities, as even a single-unit decrease holds significance in the broader context of computing resources and timing requirements for MILP solvers in full-round ciphers.

The key emphasis for differential and impossible differential cryptanalysis lies in identifying difference propagations through a cipher path. Consequently, there is a notable focus among cryptographers on developing automatic search algorithms~\cite{DBLP:conf/eurocrypt/Matsui_Algo,autosearch-sasaki,autosearch-wu,u-method,uid-method,Sasaki_IMP} tailored for the identification of these characteristics and approximations. In literature, limited efficient tools are available by an attacker who can choose distinguishable customizable parameters like primitive components of a round function or a designer can verify during construction. Also, different search options are needed for choosing input and output differences during the attack.

\subsection{Related Work}
 Mouha et al.~\cite{Mouha} were the first to articulate the challenge of determining the minimum number of active SBoxes amenable to MILP modeling for evaluating word-oriented ciphers. However, there exist ciphers that do not adhere to the word-oriented paradigm, such as PRESENT, which employs a 4-bit SBox and subsequently redistributes four bits from one SBox to four distinct SBoxes using bit-permutation.

Sun et al.~\cite{Sun_asiacrypt} devised a method to simulate all potential differential propagations bit by bit, even within the SBox, enabling the application of MILP to such structures. They leveraged MILP-based techniques to evaluate the security of block ciphers against related-key differential attacks. These methods are predominantly employed to search for single-key or related-key differential characteristics on PRESENT80, LBlock, SIMON48, DESL, and PRESENT128. Sun et al.~\cite{Sun_asiacrypt} detail various approaches for modeling the differential characteristics of an SBox using linear inequalities. 

In one approach, inequalities are derived based on specific conditional differential characteristics of an SBox. Another method involves extracting inequalities from the H-representation of the convex hull for all potential differential patterns of the SBox. To select a specific number of inequalities from the convex hull, they devised a greedy algorithm for the latter technique. They advocate for an automated method to assess the security of bit-oriented block ciphers against differential attacks, proposing numerous strategies to establish tighter security bounds using these inequalities in conjunction with MILP methodology.

Sun et al.~\cite{sun_towards} further explore the examination of differential and linear characteristics across various block ciphers using mixed-integer linear programming (MILP). They stress that a concise set of linear inequalities can effectively characterize the differential behavior of each SBox. In constructing MILP models for a variety of ciphers, Sun et al.~\cite{sun_towards} define feasible regions that accurately encapsulate all valid differential and linear properties. They assert that any subset of $\{0,1\}^n\subset\mathbb{R}^n$ can be precisely delineated through linear inequalities. They introduce a method that can discern all specified features of differential and linear properties by refining Sun et al.'s heuristic approach~\cite{Sun_asiacrypt} for identifying such characteristics into a precise one based on these MILP models.
Mouha's technique~\cite{Mouha} proves unsuitable for SPN ciphers containing diffusion layers with bitwise permutations, known as S-bP structures. This issue arises from the challenge of reconciling the diffusion effect computed concurrently by the non-linear substitution layers and bitwise permutation layers. Additionally, the MILP constraints presented by Mouha lack adequacy in modeling the differential propagation of a linear diffusion layer derived from almost-MDS or non-MDS matrices. To automatically establish a lower constraint on the number of active SBoxes for block ciphers with S-bP structures, Sun et al.~\cite{Sun_SbP} expanded upon Mouha et al.'s method~\cite{Mouha} and proposed a novel strategy grounded in mixed-integer linear programming (MILP). They successfully applied this technique to PRESENT-80.

Matsui~\cite{DBLP:conf/eurocrypt/Matsui_Algo} introduced an automated search tool utilizing the branch-and-bound search algorithm, initially applied to DES for uncovering both differential characteristics and linear approximations. To automate the quest for such characteristics in ARX ciphers, Kai Fu et al.~\cite{KaiFui_SPECK} developed a MILP model. By presuming independent inputs for modular addition and independent rounds, they leveraged the differential and linear properties of modular addition. Their exploration extended to identifying differential properties and linear approximations of the Speck cipher using the novel MILP model. The differential characteristics they identified for Speck64, Speck96, and Speck128 span one, three, and five rounds, respectively, compared to prior best differential characteristics. Moreover, the differential characteristic for Speck48 exhibits greater viability. Cui et al.~\cite{Cui} introduced an innovative automatic method for searching impossible differential trails in ciphers containing SBoxes, accounting for the differential and linear features of non-linear components, such as SBoxes themselves. They expanded the tool's functionality to include modulo addition and applied it to ARX ciphers. For HIGHT, SHACAL-2, LEA, and LBlock, the tool improves upon the current best outcomes. A new SBox modeling approach capable of addressing the likelihood of differential characteristics and reflecting a condensed version of the Differential Distribution Table (DDT) of large SBoxes was presented by Ahmed Abdelkhalek et al.~\cite{Abdelkhalek_Large_SBOX}. They increased the number of rounds necessary to resist simple differential distinguishers by one round after evaluating the upper bound on SKINNY-128's differential features. Additionally, the upper bound on differential features for two AES-round based constructions was examined.
Coming to the automatic searching of impossible differential propagations, J Kim et al.~\cite{u-method} introduced the U method, which expresses these differential propagations as matrix operations. Subsequently, Y Luo et al.~\cite{uid-method} addressed some of the U method's limitations with their UID method. For a period, encryption scheme designers widely adopted the UID method for security assessments. In a broader context, generalizing impossible differential cryptanalysis as the quest for impossible differential characteristics, S Wu and M Wang~\cite{autosearch-wu} introduced an automated matrix-based approach. This approach bridges the gap between manual searches conducted via ad hoc techniques and the UID method. Furthermore, Y Sasaki and Y Todo~\cite{autosearch-sasaki} developed an automated tool based on MILP techniques to assess a cipher's security. This tool takes inputs, including the number 'r' representing rounds and a system of linear inequalities describing the cipher's non-linear and linear layers. In turn, it generates a system of linear inequalities representing 'r' rounds of the cipher. Additional constraints can be applied to search for either a differential characteristic (that minimizes the number of active SBoxes in input and output differences) or impossible differential characteristics (that occur with zero probability in input and output differences).

\subsection{Our Contribution} 
In this paper, we introduce models for valid differential trail propagation through the linear and non-linear layers\footnote{A preliminary version of this paper is presented in CANS 2023~\cite{dpal_cans}}. Applying these primitive components we designed an MILP based tool for searching differential and impossible differential propagation's through a cipher path.   
\begin{itemize}
    \item \textbf{Modeling linear and non-linear layer} We propose two new algorithms for modeling non-linear layer and a novel \exor model for linear layer.
    
    \textbf{Greedy random-tiebreaker algorithm} We introduce a novel algorithm that selects randomly from the outcomes of the greedy algorithm. Our approach enhances the threshold for the minimum number of inequalities applicable to 4-bit SBoxes of MIBS, LBlock, and Serpent compared to the current greedy algorithm~\cite{Sun_asiacrypt}.
    
    \textbf{Subset addition approach} We propose a subset-addition-based algorithm that generates new inequalities derived from the outcomes of the H-representation of the convex hull. By incorporating $k$-subset inequalities to create new inequalities, we eliminate more improbable propagations. Subsequently, we replace a subset of old inequalities with the new ones, resulting in enhancements over existing algorithms designed for 4-bit SBoxes like Minalpher, LBlock, Serpent, Prince, and Rectangle. Moreover, the subset addition algorithm is effective for 5- and 6-bit SBoxes.

For 5-bit SBoxes such as ASCON and SC2000, we refine the boundary of inequalities. In the case of 6-bit SBoxes like APN and SC2000, we reduce the number of inequalities compared to existing results. Furthermore, we significantly improve the time required to find the minimum set of inequalities compared to the approach proposed by Boura and Coggia~\cite{BouraC}.
    
    \textbf{New XOR model} We investigated a brief depiction of the linear layer, mostly utilising XOR gates. 
    Notably, our model performs better than competing models in terms of computational efficiency while still achieving a simplified form.

    \item \textbf{Automatic differential and impossible differential searching tool} We created a tool that, when given the round function specification for an SPN block cipher, generates a MILP model for a user-specified number of rounds. The model is then solved to discover a differential characteristic that minimises the amount of active SBoxes. It also searches impossible differential characteristics corresponding to impossible transitions from the input to the output.
    
\end{itemize}

\subsection{Organization of the paper}
The organization of the paper is as follows. Section~\ref{Background} explains the background of our work. We describe the greedy random tiebreaker algorithm and its results in Section~\ref{GreedyRandomTiebreaker}. In Section~\ref{SubsetAddition}, we present the subset addition approach and the corresponding implementation process with the results. A description of modeling linear layer along with a new XOR model is provided in Section~\ref{sec:Model_Linear_Primitives}. Section~\ref{MILP_Cipher} presents the automatic tool for finding differential and impossible differential propagations.
Section~\ref{Conclusion} concludes the papaer.

\section{Background}\label{Background}
In this section we describe about the MILP modeling of the word oriented and the bit oriented block ciphers.

\subsection{Modeling word oriented block ciphers}
In Mouha et al.'s approach, word level variations propagating through a block cipher are represented as binary variables, which are subject to restrictions imposed by the word-oriented operations, XOR, nonlinear transformation and linear transformation with a predefined branch number.
Let a string P consisting of n words $P=\{P_0, P_1, \ldots, P_{n-1}\}$. We can define the difference vector $D=\{D_0, D_1, \ldots, D_{n-1}\}$ corresponding to P as follows,
\begin{equation}
  D_i=\begin{cases}
    0, & \text{if $P_i=0$}.\\
    1, & \text{otherwise}.
  \end{cases}
\end{equation}

Now taking the words of difference vector D as input or output we formalize the objective functions and the constraints of the MILP model.  
\subsubsection*{Objective Function}
Set up the objective function to be the sum of all variables representing the input words of the Sboxes.
Obviously, this objective function corresponds to the number of active Sboxes, and can be minimized to determine its lower bound. Suppose a binary variable $S_j$ represents the inputs of the \sbx's used in the encryption function, then the objective function is,
\begin{equation}
    minimize \sum_j S_j
\end{equation}

\subsubsection*{Defining Constraints}
\begin{itemize}
    \item \textbf{\exor Operation} 
    Consider an XOR gate with inputs $u_1$ and $v_1$ in one instance, and $u_2$ and $v_2$ in another. Let the corresponding outputs be $y_1$, $y_2$ and the input differences be,
\begin{flalign*}
  & \begin{aligned} & \begin{cases}
  u_1\oplus v_1&=y_1\\
    u_2\oplus v_2&=y_2
  \end{cases}
  \MoveEqLeft[-1]
  \end{aligned}
  & &
  \begin{aligned}
      & \begin{cases}
   u&=u_1\oplus u_2\\
    v&=v_1\oplus v_2
    \end{cases}
    \MoveEqLeft[-1]
  \end{aligned}
\end{flalign*}

    Suppose the output difference be $y = u \oplus v$, where $u_1,u_2,u, v_1,v_2,v, y_1,y_2,y \in F^w_2$ (w is the word size).
    The following constraints will make sure that when u, v, and y are not all zero, then there are at least two of them are nonzero:
    \begin{equation}
    \label{eq:mouha_exor}
    \centering
    \begin{cases}
    u + v + y \geq 2d,\\
    d \geq u,\\
    d \geq v,\\
    d \geq y        
    \end{cases}
    \end{equation}
    where d is a dummy variable taking values from \{0, 1\}. If each one of u, v, and y represents one bit, we
    should also add the inequality $u + v + y \leq 2$.
    
    \item \textbf{Linear Transformations}
    Let $u_{i_k}$ and $v_{j_k}$ , $k \in {0, 1, \ldots , m-1}$, be binary variables denoting the word-level input and output differences of the linear transformation L respectively.
     Since for nonzero input differences, there are totally at least $B_L$ nonzero w-bit words in the input and output differences, we include the following constraints:
    \begin{equation}
    \sum_{k=0}^{m-1} (u_{i_k} + v_{j_k}) \geq B_Ld_L
     \end{equation}
     Here $d_L \geq u_{ik}$ , $k \in \{0,\ldots,m-1\}$ and $d_L \geq v_{ik}$ , $k \in \{0,\ldots,m-1\}$  where $d_L$ is a dummy variable taking values in {0, 1} and $B_L$ is the branch number of the linear transformation.
     \item \textbf{Non-zero \sbx} In order to prevent the trivial solution, where the minimal number of active S-boxes is zero, an additional linear equation is introduced to guarantee that at least one \sbx is active.
\end{itemize}

\subsubsection*{Security evaluation}
Solve the MILP model using any MILP optimizer, and the optimized solution, say N, is the minimum number of the active Sboxes.
The probability of the best differential characteristic is upper bounded by $\epsilon^ N$, where $\epsilon$ is the maximum differential probability (MDP) of a single \sbx.

\subsubsection*{Example}
The number of active \sbx'es for AES are provided in Table~\ref{tab:active_sboxes_aes}. The MDP of AES \sbx is $2^{-6}$. So, the upper bound of the differential characteristics for eight rounds of AES is $2^{-300}$.

\begin{table}[htbp]
\centering
\caption{Minimum number of active SBoxes for AES}
\label{tab:active_sboxes_aes}
\begin{tabular}{|c|c|c|c|c|c|c|c|c|c|c|c|c|c|c|}
\hline
Rounds        & 1 & 2 & 3 & 4  & 5  & 6  & 7  & 8  & 9  & 10 & 11 & 12 & 13 & 14 \\ \hline
Active SBoxes & 1 & 5 & 9 & 25 & 26 & 30 & 34 & 50 & 51 & 55 & 59 & 75 & 76 & 80 \\ \hline
\end{tabular}
\end{table}

\subsection{Framework for Bit Oriented Ciphers}
Bit-oriented ciphers require mainly two types of constraints, constraints imposed by \exor operations (same as word oriented ciphers) and constraints for modeling the \sbx's.
\begin{itemize}
    \item \textbf{Bit level representations}
A new binary variable $x_i$ is required for every input and output bit-level difference. 
    \begin{equation*}
        x_i=\begin{cases}
            0 \textit{ if the bit difference is zero}\\
            1 \textit{ if the bit difference is one}
        \end{cases}
    \end{equation*}
Introduce another new binary variable $S_j$ for every inputs of the \sbx's used in the encryption function and the key-schedule procedure.
    \begin{equation*}
    \centering
        S_j=\begin{cases}
        1 \textit{ if the input word of the \sbx is nonzero}\\
        0 \textit{ if the input word of the \sbx is zero}
        \end{cases}
    \end{equation*}
\item \textbf{Constraints for \sbx Operations}    
 Suppose an $q \times r$ \sbx denoted by $S_t$ and the bit-level input and output differences are ($x_{i_0}, x_{i_1}, \ldots, x_{i_{q-1}}$) and ($y_{j_0}, y_{j_1}, \ldots, y_{j_{r-1}}$).
     Here we want to maintain
    \begin{equation*}
    S_t=\begin{cases}
     1 \textit{ iff at least one of } {x_{i_0},x_{i_1},\ldots,x_{i_{q-1}}} \textit{ is active}\\
    0 \textit{ iff all are zero}
    \end{cases}    
    \end{equation*}
     Equivalently, add the the following set of inequalities,
    \begin{equation}
        \begin{cases}
        x_{i_0}+x_{i_1}+\ldots+x_{i_{q-1}} - S_t\geq 0\\
        S_t-x_{i_k}>=0, \\ 
        \text{where }k\in\{0,\ldots (q-1)\}
        \end{cases}
    \end{equation}
   
For bijective Sbox's nonzero input differences must results non zero output differences and the opposite is also true,
    \begin{equation}
        \begin{cases}
        qy_{j_0}+qy_{j_1}+\ldots+qy_{j_{(r-1)}} - (x_{i_0}+ x_{i_1}+\ldots+x_{i_{q-1}}) \geq 0\\
        rx_{i_0}+ rx_{i_1}+\ldots+rx_{i_{q-1}} -(y_{j_0}+y_{j_1}+\ldots+y_{j_{(r-1)}}) \geq 0
        \end{cases}
    \end{equation}
    
The hamming weight of $(q+r)$ bit word $x_{i_0}\ldots x_{i_{q-1}} y_{j_0}\ldots y_{j_{r-1}}$
is lower bounded by branch number $B_s$ of the \sbx for non-zero input difference $x_{i_0}\ldots x_{i_{q-1}}$ and $d_s$ is a dummy variable. 
    \begin{equation}
    \begin{cases}
    \sum_{k=0}^{q-1} x_{i_k} + \sum_{t=0}^{r-1} y_{j_t} \geq B_sd_s\\
    d_s \geq x_{i_k},\\ d_s \geq y_{i_k} 
    \end{cases}
 \end{equation}
 
     where $d_s$ is a dummy variable taking values in $\{0, 1\}$ and $B_s$ is the branch number of the linear transformation.
    
\item \textbf{Updated objective function}
The new objective function is,
    \begin{equation}
        f=\textit{Minimize}\sum_j{S_j}
    \end{equation} 
 The purpose is to find the lower bound of the number of active \sbx's.   

\end{itemize}

In this section we describe the earlier used methods and algorithms for modeling SBoxes using inequalities.

\subsection{Representation of \sbx using inequalities}
An SBox $S$ can be represented as $S:\mathbb{F}_2^n \rightarrow \mathbb{F}_2^n$. We can symbolize any operation on an \sbx as $x \rightarrow y$ with $x, y \in \mathbb{F}_2^n$.  
 Let $(x_0,\ldots, x_{n-1}, y_0,\ldots, y_{n-1}) \in \mathbb{R}^{2n}$ be a $2n$-dimensional vector, where $\mathbb{R}$ is the real number field, and for an \sbx the input-output differential pattern is denoted using a point $(x_0,\ldots, x_{n-1}, y_0,\ldots, y_{n-1})$.



\subsubsection*{H-representation of the convex hull}
The convex hull of a set $P$ comprising distance points in $\mathbb{R}^n$ represents the smallest convex set that encompasses $P$. We determine the H-representation of the convex hull encompassing all possible input-output differential patterns of the \sbx by computing the Differential Distribution Table (DDT) and utilizing SageMath~\cite{sagemath} to compute the H-representation.

The H-representation furnishes $w$ linear inequalities, expressed as:

\[
A
\begin{bmatrix}
x_0\\
\vdots\\
x_{n-1}\\
y_0\\
\vdots\\
y_{n-1}
\end{bmatrix}
\leq b
\]

where $A$ represents a $w \times 2n$ matrix and both $A$ and $b$ comprise solely integer values. Each linear inequality invalidates specific points associated with impossible differential propagations. Moreover, the H-representation includes redundant inequalities linked to the MILP-based differential trail search due to the confinement of feasible points to $\{0, 1\}^{2n}$ rather than $\mathbb{R}^{2n}$. Consequently, numerous extra linear inequalities impede the efficiency of the MILP solver. Researchers employ various techniques to eliminate these redundant inequalities.

\begin{algorithm}
    Input:\\ 
    \textit{HI}: Inequalities in the H-Representation of the convex hull of an SBox.\\
    \textit{ID}: The set of all impossible differential patterns of an SBox 
    \\
    Output:\\
    $RI$: Set of inequalities that generates a stricter feasible region after maximizing the removed impossible differential patterns. 
	\begin{algorithmic}[1]
   \State $l \gets \phi, \textit{RI} \gets \phi$
    \While {\textit{ID} $\ne \phi$}
    \State $l \gets $ The inequality in \textit{HI} which maximizes the number of removed impossible differential patterns from \textit{ID}.
    \State \textit{ID} $\gets$ \textit{ID} $-$ \{\text{Removed impossible differential patterns using $l$\}}
    \State \textit{HI} $\gets$ \textit{HI} $-$ \{l\}
    \State \textit{RI} $\gets$ \textit{RI} $\cup$ \{l\}
    \EndWhile
    \State \textbf{return} \textit{RI}
    \end{algorithmic}
    \caption{Greedy Based Approach~\cite{Sun_asiacrypt}}
	\label{alg:greedy_sun}
	\end{algorithm}

\subsubsection*{Conditional differential characteristics modeling}
The logical condition that $(x_0,\ldots,x_{m-1})=(\delta_0,\ldots,\delta_{m-1}) \in \{0,1\}^m \subseteq \mathbb{Z}^m$ implies $y=\delta \in \{0,1\}\subseteq \mathbb{Z}$ can be modeled by using the following linear inequality,
        \begin{equation}        
                \sum_{i=0}^{m-1}(-1)^{\delta_i}x_i+ (-1)^{\delta+1}y-\delta+\sum_{i=0}^{m-1}\delta_i \geq 0
        \end{equation}
        
\subsubsection*{Example}        
Let $(x_0, x_1, x_2, x_3)$ and $(y_0, y_1, y_2, y_3)$ be MILP variables for the input and output differences of a 4-bit SBox. Let $(1010) \rightarrow (0111)$ be an impossible propagation in the DDT corresponding to the SBox. That is, the input difference $(1010)$ is not propagating to $(0111)$. The linear inequality which eliminates the impossible point $(1001, 0111)$ is, $-x_0 + x_1 - x_2 + x_3 + y_0 - y_1 - y_2 - y_3 + 4 \geq 0$. 
Corresponding to an SBox, if in a DDT there occur $n$ impossible paths, then at most $n$ linear inequalities are needed to model the DDT correctly. But one can reduce the value of $n$ by merging one or more available inequalities and therefore generating new inequalities.
For example if we consider two impossible propagations $(1010) \rightarrow (0111)$ and $(1010) \rightarrow (0110)$, then the linear inequality:
$-x_0 + x_1 - x_2 + x_3 + y_0 - y_1 - y_2 + y_3 \geq -3$
eliminates both the impossible points together. 


\subsection{Choosing the best inequalities from the convex hull}
Utilizing an MILP model does not assure the creation of a valid differential path. Our objective is to minimize the number of active SBoxes across a wider region, ensuring that the optimal value remains equal to or less than the minimum number of active SBoxes. Therefore, we aim to identify linear inequalities capable of trimming the MILP model while preserving the validity of the region's differential characteristics. Various algorithms have been proposed by researchers to diminish the number of inequalities in an SBox representation.

\begin{algorithm}[htbp]
    Inputs:\\ 
    $P$: Input set corresponding to the possible transitions in an SBox.\\
    $k$: The number of inequalities to be added together.
    \\
    Output:\\
    $RI$: Set of inequalities that generates more stricter feasible region after maximizing the removed impossible differential patterns. 
	\begin{algorithmic}[1]
    \State $H \gets \textit{ConvHull}(P)$
    \State \textit{RI} $\gets H$
    \For {$\textbf{all } p \in P$}
    \State Choose $k$ inequalities such that $p$ belongs to the hyperplanes of $Q_1,Q_2,\ldots,Q_k$
    \State $Q_{new}=Q_1+\ldots+Q_k$
    \State \textbf{if } $Q_{new}$ removes new impossible transitions
    \State \textit{RI} $\gets$ \textit{RI} $\cup\{Q_{new}\}$
    \State \textbf{end if}
    \EndFor
    \State \textbf{return \textit{RI}}
    \end{algorithmic}
    \caption{Modeling by selecting random set of inequalities~\cite{BouraC}}
	\label{alg:boura_select_any}
	\end{algorithm}

\subsubsection*{Greedy Algorithm Based Modeling~\cite{Sun_asiacrypt}}
The discrete points within the H-Representation (convex hull) yield numerous inequalities. An effective strategy involves identifying the most advantageous valid inequalities, which optimize the elimination of infeasible differential patterns within the convex hull's feasible region. Algorithm ~\ref{alg:greedy_sun} elucidates the greedy approach proposed by Sun et al.
\begin{algorithm}[htbp]
    Inputs:\\ 
    \textit{IDP}: Impossible differential patterns corresponding to the impossible transitions from the DDT of an SBox.\\
    $P$: Input set corresponding to the possible transitions in an SBox
    \\
    Output:\\
    $I_{SR}$: Set of inequalities that generates more stricter feasible region after maximizing the removed impossible differential patterns. 
	\begin{algorithmic}[1]
    \State H $\gets \textit{ConvHull(P)}$
    \State \textit{RI} $\gets H$
    \State Create a table \textit{PIT} of size  $|\textit{RI}| \times |\textit{IDP}|$ where $\textit{PIT}_{i,j} = 1$ if inequality \textit{$RI_i$} removes pattern $\textit{IDP}_j$, else set $\textit{PIT}_{i,j} = 0$
    \State Set $m= |\textit{RI}|$ 
    \State Generate $m$ binary variables such that $c_1, c_2, \ldots , c_m$ such that $c_i = 1$ means inequality $i$ is included in the solution space else $c_i = 0$.
   \State \textit{constraintSet}$\leftarrow \phi$
    \For {$\textbf{each } point \in \textit{IDP}$}
    \State \textit{constraintSet }$\leftarrow \textit{ constraintSet }  \cup  $ \{Construct a constraint $c_k,\ldots, c_l >= 1$ such that $p$ is removed by at least one inequality applying table \textit{PIT}\} \Comment{Generate constraints}
    \EndFor
    \State Create an objective function $\sum_{i=1}^m c_i$ with constraints \textit{constraintSet} and solve to get best inequalities ($I_{SR}$) that generate a stricter feasible region after maximizing the removed impossible differential patterns. 
    \State \textbf{return $I_{SR}$}
    \end{algorithmic}
    \caption{MILP based reduction~\cite{Sasaki_MILP_ALG}}
	\label{alg:sasaki_MILP}
	\end{algorithm}

\subsubsection*{Modeling by selecting random set of inequalities~\cite{BouraC}}
A larger collection of newly generated inequalities (resulting from the random addition of $k$ inequalities) tends to be ineffective, as they are likely to encompass the entire space $\{0,1\}^m$. When $k$ hyperplanes of the H-representation converge at a vertex within the cube $\{0,1\}^m$, aggregating corresponding inequalities may yield a novel inequality $Q_{new}$. However, the hyperplane represented by $Q_{new}$ must intersect the cube at least once. In such instances, $Q_{new}$ potentially invalidates a distinct set of infeasible points within the H-representation compared to the previous inequalities. Algorithm~\ref{alg:boura_select_any} outlines the complete procedure.
\begin{figure}[htbp]
    \centering
    \begin{tikzpicture}
        \node(start)[rect]{$\{1,2\},\{1,2,3\},\{2,3,5\},\{4,5\},\{6\},\{6,7\}$};

        \node(g1)[rect, below left=8mm and 5mm of start.south]{$\{\textcolor{mygrey}{1},\textcolor{mygrey}{2}\},\{\textcolor{mygrey}{2},\textcolor{mygrey}{3},5\},\{4,5\},\{6\},\{6,7\}$};
        \node(g2)[rect, below=8mm of g1]{$\{\textcolor{mygrey}{1},\textcolor{mygrey}{2}\},\{\textcolor{mygrey}{2},\textcolor{mygrey}{3},\textcolor{mygrey}{5}\},\{6\},\{6,7\}$};
        \node(g3)[rect, below=8mm of g2]{$\{\textcolor{mygrey}{1},\textcolor{mygrey}{2}\},\{\textcolor{mygrey}{2},\textcolor{mygrey}{3},\textcolor{mygrey}{5}\},\{\textcolor{mygrey}{6}\}$};
        \path[arrow](start.south)--node[midway, left, align=center]{Choose $\{1,2,3\}$\hspace{5mm}}(g1.north);
        \path[arrow](g1)--node[midway, left, align=center]{Choose $\{4,5\}$}(g2);
        \path[arrow](g2)--node[midway, left, align=center]{Choose $\{6,7\}$}(g3);

        \node(h1)[rect, below right=8mm and 5mm of start.south]{$\{1,\textcolor{mygrey}{2}\},\{1,\textcolor{mygrey}{2},\textcolor{mygrey}{3}\},\{4,\textcolor{mygrey}{5}\},\{6\},\{6,7\}$};
        \node(h2)[rect, below=8mm of h1]{$\{1,\textcolor{mygrey}{2}\},\{1,\textcolor{mygrey}{2},\textcolor{mygrey}{3}\},\{4,\textcolor{mygrey}{5}\},\{\textcolor{mygrey}{6}\}$};
        \node(h3)[rect, below=8mm of h2]{$\{1,\textcolor{mygrey}{2}\},\{1,\textcolor{mygrey}{2},\textcolor{mygrey}{3}\},\{\textcolor{mygrey}{6}\}$};
        \node(h4)[rect, below=8mm of h3]{$\{\textcolor{mygrey}{1},\textcolor{mygrey}{2},\textcolor{mygrey}{3}\},\{\textcolor{mygrey}{6}\}$};
        \path[arrow](start.south)--node[midway, right, align=center]{\hspace{5mm}Choose $\{2,3,5\}$}(h1.north);
        \path[arrow](h1)--node[midway, right, align=center]{Choose $\{6,7\}$}(h2);
        \path[arrow](h2)--node[midway, right, align=center]{Choose $\{4,5\}$}(h3);
        \path[arrow](h3)--node[midway, right, align=center]{Choose $\{1,2\}$}(h4);
    \end{tikzpicture}
    \caption{Variation of the output of the greedy algorithm because of a random tiebreaker}
    \label{fig:greedy_example}
\end{figure}
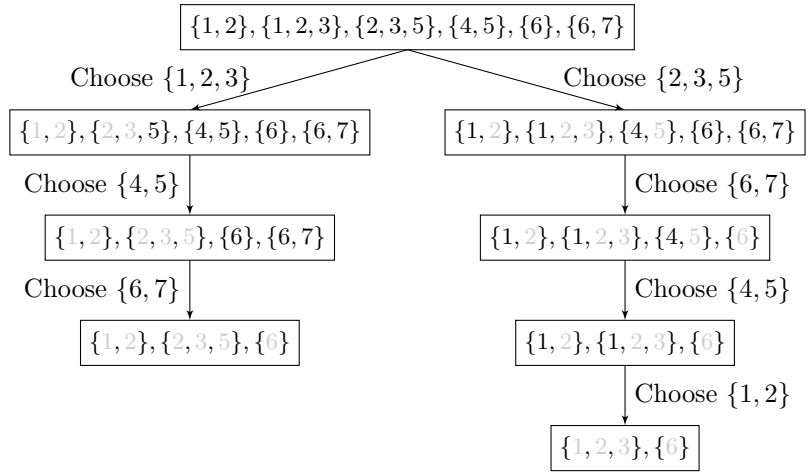

 \begin{algorithm}
    Input:\\ 
    $H_{Rep}$: Inequalities in the H-Representation of the convex hull of an SBox.\\
    $I_D$: The set of all impossible differential paths of an SBox 
    \\
    Output:\\
    $I_{SR}$: Set of n-best inequalities that generates more stricter feasible region after maximizing the removed impossible differential paths. 
	\begin{algorithmic}[1]
   \State $I_M \gets \phi, I_{SR} \gets \phi$
   \While {$I_D \ne \phi$}
    \State $I_M \gets $The inequalities in $H_{Rep}$ which maximizes the number of removed impossible differential paths from $I_D$.
    \State $I_D \gets$ $I_D-\{\text{Removed impossible differential paths using $I_M$\}}$
    \If {$Degree(I_M) > 1$}\Comment{Returns the number of elements in a set}
    \State $Rand_{I} \leftarrow \textit{ChooseRandomInequality}(I_M)$\Comment{Chooses randomly an element from a set}
    \EndIf
    \State $H_{Rep} \gets H_{Rep} - Rand_I$
    \State $I_{SR} \gets I_{SR} ~\displaystyle{\cup}~ Rand_I$
    \EndWhile
    \State \textbf{return $I_{SR}$}
    \end{algorithmic}
    \caption{Randomly select inequalities from greedy set (Greedy Random-Tiebreaker)}
	\label{alg:greedy_random}
	\end{algorithm}

\subsubsection*{MILP-based reduction algorithm}
Sasaki and Todo\cite{Sasaki_MILP_ALG} introduce a methodology that formulates a minimization problem, subsequently solved using a standard MILP solver to obtain a condensed set of inequalities. Initially, they identify all impossible differential points derived from the DDT of an SBox and derive impossible patterns based on these points. Then, for each impossible pattern, they determine which subset of inequalities renders the pattern invalid. Subsequently, they establish a constraint: every impossible pattern must be invalidated by at least one inequality within the feasible region. They construct an MILP problem aimed at minimizing the overall set of inequalities while adhering to the constraints, which they solve to obtain the minimized set of inequalities. Algorithm~\ref{alg:sasaki_MILP} delineates the entire process in a systematic manner.

\begin{algorithm}[htbp]
    Input:\\ 
    $H_{Rep}$: Inequalities in the H-Representation of the convex hull of an SBox.\\
    $I_P$: The set of all possible differential paths of an SBox 
    \\
    $I_D$: The set of all impossible differential paths of an SBox 
    \\
    Output:\\
    $I_{SR}$: Set of n-best inequalities that generates more stricter feasible region after maximizing the removed impossible differential paths. 
	\begin{algorithmic}[1]
   \State $I_{SR} \gets H_{Rep}$
   \For {$p \in I_P $}
    \State $I_H \leftarrow$ all hyperplanes in $H_{Rep}$ which the point $p$ lies on
    
    \For {$\{h_1, h_2,\ldots, h_k\} \in P_{Set}(I_H)$}
    \State $h \leftarrow h_1 + h_2 + \ldots + h_k$
    \If{$h$ is a good hyperplane (If satisfies condition of Type 1 or Type 2)}
    \State $I_{SR} \leftarrow I_{SR} \cup \{h\}$
    \EndIf
    \EndFor
    \EndFor
    \State  \textbf{return} the smallest subset of $I_{SR}$ removing all paths in $I_D$.
    \end{algorithmic}
    \caption{Generates a minimal subset of inequalities eliminating all impossible differential paths (Subset Addition)}
	\label{alg:subset_addition}
	\end{algorithm}

\begin{figure}[htbp]
    \centering
    \includegraphics[height=6.9cm, keepaspectratio]{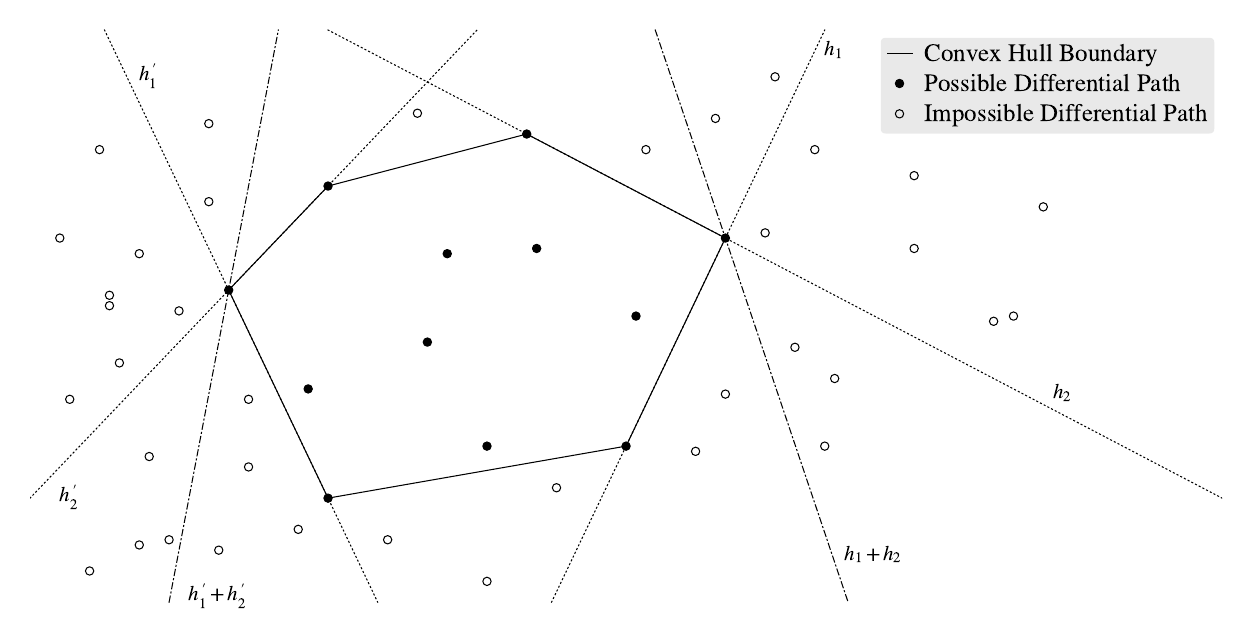}
    \caption{Separating Type 1 and Type 2 inequalities in convex hull}
    \label{fig:m_milp_diagram}
\end{figure}

\section{Modeling Non-linear Primitives: Our Approach}

\subsection{Filtering Inequalities by Greedy Random-Tiebreaker}\label{GreedyRandomTiebreaker}
We employ the original greedy algorithm proposed by S Sun et al.~\cite{Sun_asiacrypt} in our approach. Our method closely aligns with the original greedy algorithm, with the key distinction being that when multiple inequalities share the same rank, we opt for a random selection among them.

We posit that the introduction of a random tiebreaker could yield varying numbers of inequalities across multiple runs. To substantiate this claim, we illustrate it with an example of the set cover problem, which is homomorphic to the task of minimizing the MILP model of an SBox. Consider the set $S = \{1, 2, 3, 4, 5, 6, 7\}$ and its subsets $\{\{1, 2\}, \{1, 2, 3\}, \{2, 3, 5\}, \{4, 5\}, \{6\}, \{6, 7\}\}$ designed to find a cover of $S$. Two distinct greedy approaches yield covers of sizes three ($\{\{1, 2, 3\}, \{4, 5\}, \{6, 7\}\}$) and four ($\{\{2, 3, 5\}, \{4, 5\}, \{6, 7\}, \{1, 2\}\}$) (refer to Figure~\ref{fig:greedy_example}).

During numerous iterations of the greedy random-tiebreaker approach, we identify the most effective (reduced) set of inequalities capable of invalidating all impossible differential patterns. Consequently, the random tiebreaker augments the efficiency of the greedy algorithm. Table~\ref{tab:min_ineq_random_greedy} presents a comparative evaluation of the reduced number of inequalities across various 4-bit SBoxes.

The overarching methodology is encapsulated in Algorithm~\ref{alg:greedy_random}. Let $H_{Rep}$ denote the set of inequalities generated from the SageMath method $inequality\_generator()$, representing the H-representation of the convex hull of an SBox. Assume $I_D$ represents the set of all impossible differential points selected from the difference distribution table (DDT) of an SBox. Let $I_M$ store the hyperplanes in $H_{Rep}$ that eliminate the maximum number of paths from $I_D$. If $I_M$ comprises more than one element, we randomly select an inequality from $I_M$ using the \textit{ChooseRandomInequality} method and store it in $I_{SR}$. The optimal set of inequalities is retained in $I_{SR}$, and the entire process is iterated multiple times to derive the optimal set of inequalities.

\begin{table}[htbp]
    \centering
    \caption{Minimum number of inequalities for 4-bit SBoxes (Random Greedy-Tiebreaker)}
    \scalebox{0.78}{
    \begin{tabular}{ |c|c|c|c| }
        \hline
        Cipher&\thead{SageMath~\cite{sagemath}}&\thead{Sun et al.~\cite{Sun_asiacrypt}}&\thead{Random Greedy (Our Approach)}\\
        \hline
        GIFT&237&-&22\\
        KLEIN&311&22&22\\
        Lilliput&324&-&26\\
        MIBS&378&27&\bf24\\
        Midori S0&239&-&25\\
        Midori S1&367&-&24\\
        Minalpher&338&-&25\\
        Piccolo&202&23&24\\
        PRESENT&327&22&22\\
        PRIDE&194&-&22\\
        PRINCE&300&26&26\\
        RECTANGLE&267&-&23\\
        SKINNY&202&-&24\\
        TWINE&324&23&25\\
        \hline
        LBlock S0&205&28&\bf{25}\\
        LBlock S1&205&27&\bf25\\
        LBlock S2&205&27&\bf25\\
        LBlock S3&205&27&\bf26\\
        LBlock S4&205&28&\bf25\\
        LBlock S5&205&27&\bf25\\
        LBlock S6&205&27&\bf26\\
        LBlock S7&205&27&\bf25\\
        LBlock S8&205&28&\bf26\\
        LBlock S9&205&27&\bf25\\
        \hline
        Serpent S0&410&23&24\\
        Serpent S1&409&24&25\\
        Serpent S2&408&25&25\\
        Serpent S3&396&31&\bf23\\
        Serpent S4&328&26&\bf24\\
        Serpent S5&336&25&\bf23\\
        Serpent S6&382&22&\bf21\\
        Serpent S7&470&30&\bf21\\
        Serpent S8&364&-&25\\
        Serpent S9&357&-&24\\
        Serpent S10&369&-&27\\
        Serpent S11&399&-&21\\
        Serpent S12&368&-&24\\
        Serpent S13&368&-&24\\
        Serpent S14&368&-&25\\
        Serpent S15&368&-&23\\
        Serpent S16&365&-&25\\
        Serpent S17&393&-&31\\
        Serpent S18&368&-&27\\
        Serpent S19&398&-&23\\
        Serpent S20&351&-&24\\
        Serpent S21&447&-&25\\
        Serpent S22&405&-&25\\
        Serpent S23&328&-&24\\
        Serpent S24&357&-&24\\
        Serpent S25&366&-&22\\
        Serpent S26&368&-&23\\
        Serpent S27&523&-&24\\
        Serpent S28&278&-&23\\
        Serpent S29&394&-&24\\
        Serpent S30&394&-&23\\
        Serpent S31&357&-&27\\
        \hline
    \end{tabular}}
    \label{tab:min_ineq_random_greedy}
\end{table}

\subsection{Implementations and results}
We executed Algorithm~\ref{alg:greedy_random} using SageMath on a desktop computer equipped with an Intel Core i5-6500 4C/4T CPU running Manjaro Linux Sikaris 22.0.0 Xfce Edition 64-bit. During implementation, we employed a flag to randomize the list of inequalities.

\subsubsection*{Applications on 4-bit SBoxes}
We conducted our greedy random-tiebreaker algorithm on a comprehensive set of 4-bit SBoxes commonly used in ciphers. Here, we outline the comparison between our findings and the best-known results to date in terms of the minimum number of inequalities. The initial section of Table~\ref{tab:min_ineq_random_greedy} presents the outcomes for 4-bit SBoxes across 14 distinct ciphers. Our results exhibit variability; for instance, in the case of MIBS~\cite{DBLP:mibs}, we obtained 24 inequalities, three fewer than those achieved by Sun et al.~\cite{Sun_asiacrypt}. Conversely, for Prince~\cite{DBLP:prince}, our results match the existing findings, while for four other ciphers, our results are comparable.
The third segment of Table~\ref{tab:min_ineq_random_greedy} illustrates the outcomes for all the SBoxes within Serpent~\cite{DBLP:serpent}. Among the existing results for the eight SBoxes, we observe improvements for four of them (S3, S4, S5, S7), while the results remain consistent for S2. However, for the remaining three (S0, S1, S6), we experience a marginal loss, with at most a difference of two inequalities.

In the second section of Table~\ref{tab:min_ineq_random_greedy}, concerning LBlock~\cite{DBLP:lblock}, our results surpass those of Sun et al.~\cite{Sun_asiacrypt} across all SBoxes. Additionally, the reduced set of 24 inequalities for MIBS~\cite{DBLP:mibs} is detailed in Section~\ref{sec:SampleInequalities}.

\subsection{Filtering Inequalities by Subset Addition}\label{SubsetAddition}
The primary drawback of the greedy algorithm lies in its inability to guarantee an optimal solution for minimization problems. To address this limitation, we employed the Gurobi Optimizer~\cite{Gurobi}, following the methodology outlined by Sasaki and Todo~\cite{Sasaki_MILP_ALG}, to identify optimal solutions and successfully replicated their results. It's important to note that the solutions obtained using this approach represent optimal subsets of the H-representation rather than globally optimal solutions. In an attempt to overcome this, we explore the technique proposed by Boura and Coggia~\cite{BouraC}.

Boura and Coggia's algorithm focuses on generating a larger initial set of inequalities, making it easier to identify a smaller, more optimal subset. They achieve this by creating new inequalities through the addition of $k$-size subsets of existing inequalities, each representing the hyperplanes on which possible differential paths may lie. These newly generated inequalities are then evaluated based on their ability to eliminate a fresh set of impossible differential paths. However, this approach may be sluggish as it involves comparing lists of impossible differential paths removed by individual inequalities with those removed by the new inequality.

In light of this potential inefficiency, we propose an alternative algorithm that deviates from the approach outlined in~\cite{BouraC}. The core concept involves adding $k$ inequalities that represent the hyperplanes corresponding to a potential differential path ($h_1$ through $h_k$), thereby creating a novel inequality.

\begin{equation*}
    h=\sum_{i=1}^{k} h_i
\end{equation*}
and keep it only if it is good. We propose that h is good if,
\begin{itemize}
    \item \textbf{Type 1} New inequality $h$ removes more impossible differential paths than the inequality in ${h_1, h_2, h_3, \ldots, h_k}$ which removes the fewest; or
    \item \textbf{Type 2} New inequality $h$ invalidates at least as many impossible differential paths as the inequality in ${h_1, h_2, h_3, \ldots, h_k}$ which removes the most.
\end{itemize}

Algorithm~\ref{alg:subset_addition} delineates the overall process. Eventually, we determine an optimal subset using the Gurobi Optimizer. It is important to note that, unlike in~\cite{BouraC}, we do not consider which impossible differential paths are eliminated by $h$. Instead, we only assess the number of paths it removes. Figure~\ref{fig:m_milp_diagram} showcases examples of the two types of hyperplanes. These examples are purely illustrative, as the differential paths are situated on the vertices of the unit hypercube. $(h_1 + h_2)$ represents type 2, eliminating 12 impossible differential paths, whereas $h_1$ eliminates 12 and $h_2$ removes 10. Conversely, $h^{\prime}_1 + h^{\prime}_2$ belongs to type 1, removing 14 paths, while $h^{\prime}_1$ eliminates 16 and $h^{\prime}_2$ removes 10 paths respectively.

\subsubsection{Multithreading and Filtration}
Each iteration of the loop, commencing at line 4 in Algorithm \ref{alg:subset_addition}, is designed to run independently of others. Consequently, the algorithm is amenable to implementation in a multithreaded fashion utilizing a thread pool. Whenever a thread becomes available, it selects the next iteration of the loop and initiates its execution. Through this process, we observed that one particular thread consistently requires a noticeably longer time compared to others, regardless of the cipher under analysis.

This discrepancy arises because the potential differential path associated with the origin, denoted as $[0, 0, 0, 0, 0, 0, 0, 0]$, seems to intersect significantly more hyperplanes than any other path. However, it does not generate any new inequalities that contribute to the eventual optimal subset. Consequently, the thread assigned to process this specific path spends the longest duration performing non-useful tasks, rendering it apparent that this path can be disregarded from the outset.

\begin{table}[htbp]
    \centering
    \caption{Minimum number of inequalities for 4-bit SBoxes (Subset Addition)}
    \scalebox{0.81}{
    \begin{tabular}{ |c|c|c|c|c| }
        \hline
        Cipher&\thead{Sasaki and\\ Todo~\cite{Sasaki_MILP_ALG}}&\thead{Boura and\\ Coggia~\cite{BouraC}}&\thead{Subset Addition ($k = 2$)\\(Our approach)}&\thead{Subset Addition ($k = 3$)\\(Our approach)}\\
        \hline
        GIFT&-&17&17&17\\
        KLEIN&21&19&19&19\\
        Lilliput&23&19&20&19\\
        MIBS&23&20&20&20\\
        Midori S0&21&16&17&16\\
        Midori S1&22&20&20&20\\
        Minalpher&22&19&19&\bf18\\
        Piccolo&21&16&16&16\\
        PRESENT&21&17&17&17\\
        PRIDE&-&16&17&17\\
        PRINCE&22&19&19&\bf18\\
        RECTANGLE&21&17&17&\bf16\\
        SKINNY&21&16&16&16\\
        TWINE&23&19&20&19\\
        \hline
        LBlock S0&24&17&17&\bf16\\
        LBlock S1&24&17&17&\bf16\\
        LBlock S2&24&17&17&\bf16\\
        LBlock S3&24&17&17&\bf16\\
        LBlock S4&24&17&17&\bf16\\
        LBlock S5&24&17&17&\bf16\\
        LBlock S6&24&17&17&\bf16\\
        LBlock S7&24&17&17&\bf16\\
        LBlock S8&24&17&17&\bf16\\
        LBlock S9&24&17&17&\bf16\\
        \hline
        Serpent S0&21&17&18&17\\
        Serpent S1&21&17&19&18\\
        Serpent S2&21&18&18&\bf17\\
        Serpent S3&27&20&16&\bf14\\
        Serpent S4&23&19&19&19\\
        Serpent S5&23&19&17&\bf17\\
        Serpent S6&21&17&16&\bf16\\
        Serpent S7&27&20&16&\bf16\\
        Serpent S8&-&-&18&18\\
        Serpent S9&-&-&18&17\\
        Serpent S10&-&-&17&16\\
        Serpent S11&-&-&15&15\\
        Serpent S12&-&-&18&18\\
        Serpent S13&-&-&18&18\\
        Serpent S14&-&-&18&18\\
        Serpent S15&-&-&18&18\\
        Serpent S16&-&-&17&16\\
        Serpent S17&-&-&19&19\\
        Serpent S18&-&-&18&18\\
        Serpent S19&-&-&18&17\\
        Serpent S20&-&-&19&19\\
        Serpent S21&-&-&18&17\\
        Serpent S22&-&-&17&16\\
        Serpent S23&-&-&19&19\\
        Serpent S24&-&-&18&17\\
        Serpent S25&-&-&17&16\\
        Serpent S26&-&-&18&18\\
        Serpent S27&-&-&17&16\\
        Serpent S28&-&-&17&17\\
        Serpent S29&-&-&17&17\\
        Serpent S30&-&-&17&17\\
        Serpent S31&-&-&18&17\\
        \hline
    \end{tabular}}
    \label{tab:min_ineq_subsetaddition}
\end{table}

\subsubsection{Implementation and Results}
We implemented Algorithm~\ref{alg:subset_addition} in C++ on a desktop computer equipped with an Intel Core i5-6500 4C/4T CPU running Manjaro Linux Sikaris 22.0.0 Xfce Edition 64-bit. Subsequently, we extended the program to incorporate the Gurobi Optimizer for identifying the optimal subset of inequalities. Users are provided with the option to choose (by defining a macro during compilation) between good hyperplanes of types 1 and 2. Our experiments indicate that selecting type 1 is either superior to or on par with selecting type 2.
\subsubsection*{Application to 4-bit SBoxes}
Algorithm~\ref{alg:subset_addition} has been effectively applied to most 4-bit SBoxes. Among the 14 4-bit SBoxes listed in the first section of Table~\ref{tab:min_ineq_subsetaddition}, we observe improved results for Prince~\cite{DBLP:prince}, Minalpher~\cite{Minalpher}, and Rectangle~\cite{DBLP:rectangle} when setting $k=3$. However, the results remain consistent with existing ones for 10 SBoxes. Only for PRIDE~\cite{PRIDE}, do we note an additional minimum number of inequalities.

For all ten LBlock~\cite{DBLP:lblock} SBoxes, the inequality count decreases to 16 from 17~\cite{BouraC}. The outcomes for LBlock are detailed in the second section of Table~\ref{tab:min_ineq_subsetaddition} for both $k=2$ and $k=3$ settings.

The third section of Table~\ref{tab:min_ineq_subsetaddition} illustrates the results for 32 Serpent~\cite{DBLP:serpent} SBoxes. Comparing with the existing results of eight Serpent SBoxes $(S0$ to $S7)$, we improve the results for five $(S2, S3, S5, S6, S7)$. However, for two SBoxes $(S0, S4)$, the results remain unchanged, while for $S2$, we observe a decrease. For the remaining 24 SBoxes, we present new results. The inequalities for Serpent S3 are detailed in Section~\ref{sec:SampleInequalities}.   

\subsubsection*{Application to 5- and 6-bit SBoxes}
We applied Algorithm~\ref{alg:subset_addition} to ASCON~\cite{DBLP:ascon5} and SC2000~\cite{DBLP:sc2000-5}, which utilize 5-bit SBoxes. In this scenario, by setting $k=3$, we observed improved results. For 6-bit SBoxes of APN~\cite{APN} and SC2000 with $k=2$, we surpassed the existing boundary set by Boura and Coggia~\cite{BouraC}. However, for the 5-bit SBox of FIDES~\cite{FIDES}, we obtained one additional inequality compared to the existing boundary. The results for 5 and 6-bit SBoxes are presented in Table~\ref{tab:min_ineq_5_6_bit_sbox}, alongside existing results, for comparison.

\subsubsection*{Reducing running time over Boura and Coggia~\cite{BouraC} technique}
As previously noted, given that each impossible differential path is handled independently, parallel processing can be utilized to decrease the overall running time. In this approach, a worker thread can independently process a distinct possible differential path at a time.

\begin{table}[htbp]
\centering
\caption{ Minimum number of Inequalities for 5- and 6-bit SBoxes}
\label{tab:min_ineq_5_6_bit_sbox}
\scalebox{0.93}{
\begin{tabular}{|c|c|c|c|c|c|}
\hline
SBox   & \thead{SBox\\ Size} & SageMath~\cite{sagemath} & Boura and Coggia~\cite{BouraC} & \thead{Subset Addition\\ $k=2$\\(Our approach)} & \thead{Subset Addition\\ $k=3$\\(Our approach)} \\ \hline
ASCON  & \multirow{3}{*}{5}                                  & 2415     & 32           & \textbf{31}                                                   & \textbf{31}                                                    \\ \cline{1-1} \cline{3-6} 
FIDES  &                                                     & 910      & 61           & 64                                                   & 62                                                    \\ \cline{1-1} \cline{3-6} 
SC2000 &                                                     & 908      & 64           & 65                                                   & \textbf{63}                                                    \\ \hline
APN    & \multirow{3}{*}{6}                                  & 5481     & 167          & \textbf{163}                                                  & \multirow{3}{*}{--}                                   \\ \cline{1-1} \cline{3-5}
FIDES  &                                                     & 7403     & 180          & 184                                                  &                                                       \\ \cline{1-1} \cline{3-5}
SC2000 &                                                     & 11920    & 214          & \textbf{189}                                                  &                                                       \\ \hline
\end{tabular}}
\end{table}

To implement multithreading, we utilized the C++ POSIX Threads API, encapsulated within the thread library. In Table~\ref{table:m_milp_time}, we present the runtime data of our algorithm (Algorithm ~\ref{alg:subset_addition}) for several SBoxes. We report the average running times for LBlock S0 through S9 and Serpent S0 through S7. Generally, it seems that larger values of $k$ tend to result in smaller subsets of inequalities. However, we were unable to confirm this hypothesis definitively. While in our experiments, $k = 3$ typically produced smaller subsets than $k = 2$, testing with $k = 4$ posed challenges. For Lilliput~\cite{Lilliput}, MIBS~\cite{DBLP:mibs}, and Serpent S3, the outputs did not improve with $k = 4$, but the memory requirement soared to around 10 GiB. Consequently, we couldn't test other ciphers due to these limitations.

Boura and Coggia~\cite{BouraC} reported that their algorithm implementation took a few minutes for $k = 2$, while for $k = 3$, it required a few hours. They haven't provided precise estimates of running times for different SBoxes using their approach. While we attempted to replicate their results, the runtime on our system exceeded our expectations. Nonetheless, we compare our algorithm for $k = 2$ and $k = 3$ with that of Boura and Coggia~\cite{BouraC}, and our results are detailed in Table~\ref{table:m_milp_time}. Our implementation is faster by two orders of magnitude and yields comparable results. We achieved significantly better running times after implementing all the optimizations mentioned earlier in our program.
\begin{table}[htbp]
\centering
\caption{Approximate running time of Subset Addition Algorithm}
\label{table:m_milp_time}
\begin{tabular}{|c|cc|}
\hline
\multirow{2}{*}{SBox} & \multicolumn{2}{c|}{Required Time for Algorithm~\ref{alg:subset_addition} (in sec)} \\ \cline{2-3} 
                      & \multicolumn{1}{c|}{No. of Inequality, k = 2} & No. of Inequality, k = 3                          \\ \hline
Klein                 & \multicolumn{1}{c|}{0.16}                           & 2.5                          \\ \hline
LBlock S*             & \multicolumn{1}{c|}{0.19}                           & 2.2                          \\ \hline
MIBS                  & \multicolumn{1}{c|}{1.9}                            & 4.5                          \\ \hline
Piccolo               & \multicolumn{1}{c|}{0.15}                           & 2.0                          \\ \hline
PRESENT               & \multicolumn{1}{c|}{0.28}                           & 3.9                          \\ \hline
PRINCE                & \multicolumn{1}{c|}{0.17}                           & 4.8                          \\ \hline
Serpent S*            & \multicolumn{1}{c|}{0.49}                           & 8.3                          \\ \hline
TWINE                 & \multicolumn{1}{c|}{0.16}                           & 3.4                          \\ \hline
\end{tabular}
\end{table}

\subsection{Sample Reduced Inequalities}\label{sec:SampleInequalities}
Applying random greedy tiebreaker algorithm~\ref{alg:greedy_random} for MIBS~\cite{DBLP:mibs}, the reduced 24 inequalities are as follows,
\begin{verbatim}
 - 1x3 - 2x2 - 2x1 - 1x0 + 4y3 + 5y2 + 5y1 + 5y0 >= 0
 + 5x3 + 4x2 + 4x1 + 3x0 - 1y3 - 2y2 + 1y1 - 2y0 >= 0
 - 2x3 + 2x2 + 4x1 + 1x0 + 3y3 + 1y2 - 3y1 - 3y0 >= -4
 - 1x3 - 4x2 + 3x1 + 2x0 - 1y3 - 3y2 + 4y1 + 2y0 >= -5
 - 2x3 + 1x2 - 3x1 - 1x0 - 1y3 - 3y2 - 2y1 - 2y0 >= -11
 - 1x3 - 2x2 - 4x1 + 4x0 - 4y3 + 2y2 + 1y1 - 3y0 >= -10
 + 2x3 - 1x2 + 3x1 + 1x0 - 2y3 + 2y2 - 3y1 + 1y0 >= -3
 + 1x3 + 2x2 - 4x1 + 2x0 + 3y3 + 1y2 + 2y1 + 4y0 >= 0
 + 1x3 + 3x2 - 2x1 - 3x0 + 1y3 + 3y2 + 2y1 - 1y0 >= -3
 + 2x3 - 1x2 - 2x1 - 2x0 - 1y3 - 1y2 - 2y1 + 0y0 >= -7
 + 0x3 + 2x2 + 2x1 - 1x0 + 1y3 + 1y2 - 1y1 + 1y0 >= 0
 - 3x3 - 3x2 + 1x1 - 2x0 + 1y3 - 2y2 + 1y1 + 2y0 >= -7
 + 2x3 - 1x2 + 2x1 - 1x0 + 1y3 + 1y2 + 2y1 - 1y0 >= -1
 + 1x3 - 2x2 - 2x1 + 2x0 + 1y3 + 1y2 - 1y1 - 2y0 >= -5
 - 1x3 + 2x2 - 1x1 + 1x0 + 2y3 - 2y2 + 1y1 - 1y0 >= -3
 - 1x3 + 1x2 + 0x1 - 1x0 - 1y3 - 1y2 + 0y1 + 1y0 >= -3
 + 1x3 - 2x2 - 1x1 - 1x0 + 1y3 - 2y2 - 2y1 + 1y0 >= -6
 + 2x3 - 1x2 + 0x1 - 2x0 - 2y3 + 2y2 - 1y1 + 1y0 >= -4
 - 1x3 - 1x2 + 1x1 - 1x0 - 1y3 + 0y2 - 1y1 - 1y0 >= -5
 - 1x3 + 1x2 - 1x1 + 2x0 + 1y3 + 2y2 - 1y1 + 2y0 >= -1
 + 2x3 + 1x2 + 2x1 + 3x0 - 2y3 - 1y2 - 1y1 + 2y0 >= -1
 - 3x3 - 2x2 + 1x1 + 3x0 - 1y3 + 1y2 + 2y1 + 3y0 >= -3
 + 1x3 - 1x2 - 2x1 - 2x0 - 1y3 - 1y2 - 1y1 - 1y0 >= -7
 - 1x3 + 1x2 + 0x1 - 1x0 - 1y3 + 1y2 + 1y1 - 1y0 >= -3
 \end{verbatim}
Applying subset addition algorithm~\ref{alg:subset_addition} for Serpent S3 the inequalities are as follows,
\begin{verbatim}
 - 5x3 + 4x2 + 4x1 - 5x0 + 2y3 + 10y2 + 3y1 + 10y0 >= 0
 + 6x3 - 1x2 - 2x1 + 2x0 + 1y3 +  7y2 - 3y1 +  7y0 >= 0
 - 2x3 + 0x2 - 3x1 - 3x0 - 2y3 -  4y2 - 1y1 +  4y0 >= -11
 + 3x3 + 0x2 + 3x1 + 2x0 + 1y3 -  4y2 + 2y1 +  4y0 >= 0
 - 3x3 - 3x2 + 0x1 - 2x0 - 1y3 +  4y2 - 2y1 -  4y0 >= -11
 - 4x3 - 4x2 - 1x1 - 3x0 + 1y3 +  2y2 - 1y1 -  4y0 >= -13
 + 2x3 - 2x2 + 1x1 - 4x0 - 4y3 +  3y2 + 2y1 -  4y0 >= -10
 + 2x3 + 6x2 + 2x1 + 1x0 - 3y3 -  4y2 - 4y1 -  4y0 >= -10
 - 2x3 + 8x2 + 4x1 - 1x0 + 5y3 -  7y2 + 6y1 -  7y0 >= -10
 - 2x3 - 5x2 - 1x1 + 2x0 - 3y3 -  5y2 + 3y1 -  5y0 >= -17
 + 2x3 + 3x2 + 0x1 + 3x0 + 2y3 +  4y2 + 1y1 -  4y0 >= 0
 + 4x3 - 3x2 - 2x1 + 0x0 + 2y3 -  3y2 - 1y1 -  3y0 >= -9
 - 2x3 - 1x2 + 2x1 + 4x0 + 4y3 -  4y2 - 2y1 +  3y0 >= -5
 + 0x3 - 1x2 - 1x1 + 5x0 - 2y3 +  5y2 + 2y1 +  5y0 >= 0
 \end{verbatim}

Applying subset addition algorithm~\ref{alg:subset_addition} for ASCON SBox the 31 inequalities are as follows,
\begin{verbatim}

 - 9x5 +  8x4 +  6x3 + 11x2 - 6x1 + 4y5 - 5y4 - 3y3 - 1y2 +  3y1 >=  12
 - 1x5 +  5x4 +  8x3 +  7x2 - 3x1 + 8y5 + 7y4 - 2y3 + 1y2 -  2y1 >=  0
 + 1x5 +  2x4 +  4x3 +  2x2 - 2x1 - 4y5 - 3y4 - 2y3 + 0y2 -  4y1 >=  11
 + 5x5 + 11x4 +  4x3 + 11x2 + 6x1 - 3y5 + 2y4 - 1y3 + 0y2 -  7y1 >=  0
 + 5x5 +  7x4 -  6x3 +  3x2 - 3x1 + 6y5 - 1y4 - 1y3 + 4y2 +  1y1 >=  4
 - 1x5 +  7x4 +  7x3 +  9x2 - 3x1 + 9y5 - 3y4 - 2y3 - 1y2 +  9y1 >=  0
 - 1x5 -  2x4 +  0x3 +  2x2 - 1x1 - 3y5 - 3y4 + 2y3 + 1y2 +  2y1 >=  7
 - 1x5 +  7x4 +  9x3 +  8x2 - 3x1 - 3y5 + 9y4 - 2y3 - 1y2 + 10y1 >=  0
 - 2x5 +  5x4 +  2x3 -  5x2 - 3x1 - 2y5 + 0y4 + 0y3 - 1y2 -  3y1 >=  11
 + 1x5 -  2x4 +  0x3 -  1x2 + 2x1 + 0y5 + 1y4 + 2y3 - 2y2 -  2y1 >=  5
 + 2x5 -  1x4 +  0x3 +  1x2 + 2x1 + 0y5 + 0y4 - 2y3 + 2y2 -  1y1 >=  2
 + 3x5 +  2x4 +  0x3 -  3x2 + 1x1 - 2y5 + 0y4 - 1y3 - 2y2 +  3y1 >=  5
 + 3x5 +  5x4 +  4x3 -  4x2 + 2x1 + 3y5 + 0y4 - 1y3 - 1y2 +  3y1 >=  0
 - 2x5 +  0x4 -  1x3 -  2x2 - 2x1 + 1y5 + 1y4 + 3y3 + 3y2 +  0y1 >=  4
 + 2x5 -  3x4 +  3x3 -  3x2 - 2x1 + 0y5 + 0y4 + 1y3 + 3y2 +  1y1 >=  5
 + 2x5 +  2x4 -  2x3 -  2x2 - 1x1 + 0y5 + 0y4 + 0y3 + 2y2 +  1y1 >=  3
 + 1x5 -  4x4 -  4x3 +  3x2 + 2x1 + 2y5 + 2y4 - 1y3 - 1y2 +  2y1 >=  6
 - 1x5 -  3x4 - 12x3 + 10x2 + 4x1 - 9y5 + 8y4 - 5y3 + 1y2 -  7y1 >=  25
 - 3x5 -  1x4 -  6x3 +  6x2 - 1x1 - 5y5 - 5y4 - 3y3 + 2y2 +  7y1 >=  17
 + 0x5 +  3x4 -  2x3 +  3x2 + 2x1 - 1y5 - 1y4 + 2y3 + 0y2 -  1y1 >=  2
 - 1x5 +  4x4 -  2x3 + 10x2 - 6x1 + 5y5 + 5y4 + 4y3 - 1y2 +  6y1 >=  0
 - 2x5 +  2x4 -  2x3 -  2x2 + 1x1 + 2y5 + 0y4 + 0y3 + 0y2 -  1y1 >=  5
 + 6x5 -  5x4 -  6x3 -  2x2 + 3x1 - 1y5 - 6y4 + 0y3 - 3y2 +  1y1 >=  17
 + 2x5 -  2x4 -  2x3 -  3x2 - 2x1 + 0y5 + 3y4 + 1y3 + 1y2 +  1y1 >=  6
 - 2x5 -  1x4 -  1x3 -  3x2 - 3x1 + 0y5 - 1y4 - 2y3 - 2y2 +  0y1 >=  12
 + 0x5 -  2x4 -  1x3 +  2x2 + 0x1 + 2y5 - 2y4 + 1y3 + 0y2 -  2y1 >=  5
 + 0x5 -  2x4 +  3x3 +  4x2 + 3x1 - 1y5 + 1y4 + 4y3 + 0y2 -  1y1 >=  0
 - 2x5 -  2x4 +  2x3 -  1x2 + 1x1 + 0y5 - 2y4 + 1y3 + 2y2 -  2y1 >=  7
 - 2x5 -  1x4 +  2x3 -  2x2 + 2x1 + 1y5 - 2y4 + 0y3 - 2y2 +  2y1 >=  7
 - 3x5 -  3x4 -  1x3 -  1x2 - 2x1 + 0y5 + 2y4 - 4y3 - 4y2 -  3y1 >=  17
 - 2x5 -  2x4 -  1x3 -  2x2 + 2x1 + 0y5 + 2y4 + 0y3 + 1y2 +  1y1 >=  5

\end{verbatim}

\section{Modeling Linear Primitives}\label{sec:Model_Linear_Primitives}

The \exor modeling approach (see Equation~\ref{eq:mouha_exor}) proposed by Mouha et al.~\cite{Mouha} does not accurately model an \exor gate. For example, the solution 
$ u = 1, v = 1, y = 1, d = 1$ implies that if
$u_1\oplus v_1=y_1$ then $\overline{u_1}\oplus\overline{v_1}=\overline{y_1}$
where the overline denotes complementation. However, the last statement is clearly false.

A more robust model for \exor gates, which overcomes the limitations of the previous approach, was introduced by Jun Yin et al.~\cite{yin} based on the research of Yu Sasaki and Yosuke Todo~\cite{Sasaki_MILP_ALG}. This improved model, formalized by M. B. Ilter and A. A. Selçuk~\cite{xormodel-chull-kleinprince}, dispenses with the use of dummy variables and offers a more accurate representation of \exor gates. The fundamental idea behind this model is to treat an n-input \exor gate as a transformation, similar to an \sbx. By computing the convex hull of potential differential paths and applying \cite{Sasaki_MILP_ALG}'s MILP technique to find an optimal subset, this approach allows for a thorough analysis of \exor gates. For instance, the case of $n = 4$ is illustrated in Table~\ref{tab:4xor-ddt} of the reference.

\begin{table}[htb]
    \centering
     \caption{DDT of a four-input XOR gate}
    \label{tab:4xor-ddt}
    \begin{tabular}{|r|rr|}
        \hline
        &$\mathtt{0_2}$&$\mathtt{1_2}$\\
        \hline
        $\mathtt{0_{16}}$&16&0\\
        $\mathtt{1_{16}}$&0&16\\
        $\mathtt{2_{16}}$&0&16\\
        $\mathtt{3_{16}}$&16&0\\
        $\mathtt{4_{16}}$&0&16\\
        $\mathtt{5_{16}}$&16&0\\
        $\mathtt{6_{16}}$&16&0\\
        $\mathtt{7_{16}}$&0&16\\
        $\mathtt{8_{16}}$&0&16\\
        $\mathtt{9_{16}}$&16&0\\
        $\mathtt{A_{16}}$&16&0\\
        $\mathtt{B_{16}}$&0&16\\
        $\mathtt{C_{16}}$&16&0\\
        $\mathtt{D_{16}}$&0&16\\
        $\mathtt{E_{16}}$&0&16\\
        $\mathtt{F_{16}}$&16&0\\
        \hline
    \end{tabular}
   
\end{table}

As before, row indices indicate input differences, and column indices indicate output differences. 
\begin{equation*}
    \{(0,0,0,0,0), (0, 0, 0, 1, 1), (0, 0, 1, 0, 1),\ldots, (1, 1, 1, 1, 0)\}
\end{equation*}

after which, the inequalities are found just like in the case of the non-linear layer (as described in the previous
section). According to \cite{xormodel-chull-kleinprince}, an \exor gate with $n$ input bits requires $2^n$ inequalities
to model accurately.

\subsection{An Example of 3-input \exor modeling}
Let $D = A \oplus B \oplus C$ where $A,B,C,D \in \mathbb F_2$. We write the possible points corresponding to $(A, B, C, D)$ for transitions as follows,
\begin{equation*}
\{(0, 0, 0, 0), (0, 0, 1, 1), (0, 1, 1, 0), (1, 1, 0, 0), (1, 0, 1, 0), (0, 1, 0, 1), (0, 1, 1, 0), (1, 1, 1, 1)\}    
\end{equation*}

A polyhedron with an H-representation has a set of given valid points in its representation. Among 16 constraints, some of which are redundant are derived by computing the H-representation of these valid points. The goal is to avoid all impossible transitions like 
\begin{equation*}
    \{(0, 0, 0, 1), (0, 0, 1, 0), (0, 1, 0, 0), (1, 0, 0, 0), (1, 1, 1, 0), (1, 1, 0, 1), (1, 0, 1, 1), (0, 1, 1, 1)\}
\end{equation*}
Now find the fewest equations feasible to represent the 2-input \exor operation. The reduction approach developed by Sasaki and Todo~\cite{Sasaki_IMP} is used for this procedure. The eight constraints (see Table~\ref{tab:3_xor_impossible_trans}) are derived for the 3-input \exor model after guaranteeing that all impossible points are eliminated.

\begin{table}[htbp]
\centering
\caption{Inequalities corresponding to impossible transitions}
\label{tab:3_xor_impossible_trans}
\begin{tabular}{|c|c|}
\hline
Impossible Transition & Inequality                        \\ \hline
$(0, 0, 0, 1)$        & $A + B + C - D >= 0$              \\ \hline
$(0,0,1,0)$           & $A + B - C + D >= 0$              \\ \hline
$(0,1,0,0)$           & $A - B + C + D >= 0$ \\ \hline
$(1,0,0,0)$           & $-A + B + C + D >= 0$             \\ \hline
$(1,1,1,0)$           & $-A - B - C + D >= -2$            \\ \hline
$(1,1,0,1)$           & $-A - B + C - D >= -2$            \\ \hline
$(1,0,1,1)$           & $-A + B - C - D >= -2$            \\ \hline
$(0,1,1,1)$           & $A - B - C - D >= -2$             \\ \hline
\end{tabular}

\end{table}

\subsection{Representation of \mc Layer}
For \mc matrix (MDS or MDS-like), we convert it to its primitive representation. 
For example in the case of Midori, convert the \mc matrix of midori into the following $16 \times 16$ binary matrix M,
\begin{align*}
    M= \begin{bmatrix}
    0000 & 1000 & 1000 & 1000\\
    0000 & 0100 & 0100 & 0100\\
    0000 & 0010 & 0010 & 0010\\
    0000 & 0001 & 0001 & 0001\\\\
1000 & 0000 & 1000 & 1000\\
0100 & 0000 & 0100 & 0100\\
0010 & 0000 & 0010 & 0010\\
0001 & 0000 & 0001 & 0001\\\\
1000 & 1000 & 0000 & 1000\\
0100 & 0100 & 0000 & 0100\\
0010 & 0010 & 0000 & 0010\\
0001 & 0001 & 0000 & 0001\\\\
1000 & 1000 & 1000 & 0000\\
0100 & 0100 & 0100 & 0000\\
0010 & 0010 & 0010 & 0000\\
0001 & 0001 & 0001 & 0000\\
\end{bmatrix}   
\end{align*}

Now for the \mc operation of Midori64 we apply, $M \times X = Y$, where M is the $(16 \times 16)$ binary \mc matrix. X and Y is the $(16 \times 1)$ matrices, representing one column of input state and output state respectively. Hence, each output bit of Midori's \mc can be written as a form of three input XOR. 
The valid propagation can be modeled by removing eight impossible transitions for each bit. Hence, all the valid propagation patterns can be modeled with $64 \times 8 = 512$ inequalities. The number of inequalities depends on the hamming weight(HW) of any row of a \mc matrix. 
For each bit result with n variables $2^{n-1}$ inequalities are required.

\subsection{Difference between \sbx and \exor Modeling}
We briefly describe the DDT of Liliput \sbx, presented in Table~\ref{tab:lilliput-ddt} and the DDT of a four input \exor gate (see Table~\ref{tab:4xor-ddt}) to highlight significant distinctions between a conventional \sbx and an \exor gate. These distinctions extend beyond the well-known fact that \sbx'es execute non-linear transformations while \exor gates carry out linear transformations.

\begin{table}[htbp]
    \centering
       \caption{DDT of the \sbx of Lilliput}
    \label{tab:lilliput-ddt}
    \scalebox{0.90}{
    \begin{tabular}{|r|rrrrrrrrrrrrrrrr|}
        \hline
        &$\mathtt{0_{16}}$&$\mathtt{1_{16}}$&$\mathtt{2_{16}}$&$\mathtt{3_{16}}$&$\mathtt{4_{16}}$&$\mathtt{5_{16}}$&$\mathtt{6_{16}}$&$\mathtt{7_{16}}$&$\mathtt{8_{16}}$&$\mathtt{9_{16}}$&$\mathtt{A_{16}}$&$\mathtt{B_{16}}$&$\mathtt{C_{16}}$&$\mathtt{D_{16}}$&$\mathtt{E_{16}}$&$\mathtt{F_{16}}$\\
        \hline
        $\mathtt{0_{16}}$&16&0&0&0&0&0&0&0&0&0&0&0&0&0&0&0\\
        $\mathtt{1_{16}}$&0&2&0&0&0&0&2&0&0&2&2&2&4&0&0&2\\
        $\mathtt{2_{16}}$&0&0&0&2&2&0&2&2&0&4&0&2&0&2&0&0\\
        $\mathtt{3_{16}}$&0&2&0&0&0&2&2&2&2&0&0&0&0&2&0&4\\
        $\mathtt{4_{16}}$&0&0&0&2&0&2&0&0&0&0&2&4&0&2&2&2\\
        $\mathtt{5_{16}}$&0&4&2&2&0&2&0&2&0&2&2&0&0&0&0&0\\
        $\mathtt{6_{16}}$&0&0&2&0&0&0&4&2&0&0&2&0&2&2&2&0\\
        $\mathtt{7_{16}}$&0&0&0&2&2&2&2&0&2&0&4&0&2&0&0&0\\
        $\mathtt{8_{16}}$&0&2&2&4&2&0&2&0&0&0&0&0&0&0&2&2\\
        $\mathtt{9_{16}}$&0&0&0&2&0&0&0&2&4&2&0&0&2&0&2&2\\
        $\mathtt{A_{16}}$&0&0&2&0&2&0&0&4&2&0&2&2&0&0&0&2\\
        $\mathtt{B_{16}}$&0&2&0&0&2&2&0&2&0&0&0&2&2&0&4&0\\
        $\mathtt{C_{16}}$&0&2&0&0&2&0&0&0&2&2&2&0&0&4&2&0\\
        $\mathtt{D_{16}}$&0&2&4&2&0&0&0&0&2&0&0&2&2&2&0&0\\
        $\mathtt{E_{16}}$&0&0&2&0&4&2&0&0&0&2&0&0&2&2&0&2\\
        $\mathtt{F_{16}}$&0&0&2&0&0&4&2&0&2&2&0&2&0&0&2&0\\
        \hline
    \end{tabular}}
 
\end{table}
\begin{itemize}
    \item \textbf{Output Size}
 \sbx outputs are typically multi-bit, meaning that an invertible \sbx with an n-bit input will produce n-bit outputs. In contrast, \exor gate outputs are single-bit, where an n-bit input is reduced to a single bit.
 \item \textbf{Bijectivity}
 In most cases, \sbx'es are designed to be bijective, ensuring a one-to-one and invertible relationship between input and output to prevent any loss of information. Conversely, XOR gates exhibit a many-to-one relationship, making them inherently non-bijective.
\item \textbf{Difference Propagation}
Referencing Table~\ref{tab:lilliput-ddt}, it becomes evident that \sbx differences exhibit a one-to-many relationship, wherein a single input difference can map to multiple output differences. Conversely, as depicted in Table~\ref{tab:4xor-ddt}, \exor gate differences demonstrate a many-to-one relationship, indicating that several input differences can lead to the same output difference.

 \item \textbf{Output Difference Predictability}
These observations naturally stem from the earlier points. Understanding the input difference of an \sbx does not uniquely determine the corresponding output difference. In contrast, when it comes to an \exor gate, knowledge of the input difference directly determines the corresponding output difference.
\end{itemize}

The final observations hold substantial importance. The relationship between input and output differences in an \exor gate exhibits the characteristics of a well-defined function. Leveraging this insight, it becomes possible to create an efficient model for the linear layer.

\subsection{New \exor modeling}
MILP involves the optimization of linear constraints within a problem containing a mix of both integral and non-integral variables. However, when applied in the realm of differential cryptanalysis, it's important to note that all variables are artificially constrained to be binary integers, essentially transforming the problems into ILP, or Integer Linear Programming, problems.

In contrast, MILP solvers are engineered to be highly versatile, capable of handling a wide range of variable types, including non-integral, integral non-binary, and binary variables. Hence, for an $n$-input \exor gate:
\begin{align*}
    u_0=u_1\oplus u_2\oplus u_3\oplus\cdots\oplus a_n
\end{align*}
we propose the following constraint,
\begin{align}
    y_0+y_1+y_2+y_3+\cdots+y_n=2d\label{eq:xor-better-model}
\end{align}
Here, $y_0$ through $y_n$ are binary variables; $y_i$ represents the difference corresponding to $u_i$ and $u_i$. We introduce $d$, a dummy integral variable, to capture the behaviour of the \exor gate. In general, Equation~\ref{eq:xor-better-model} conveys the concept that differences can be introduced in an even number of bits. In simpler terms, the parity (odd or even) of the input difference vector must align with that of the output difference.

\begin{table}[htbp]
\centering
\caption{DDT of an $n$-input \exor gate}
    \label{tab:nxor-ddt}
\begin{tabular}{|cc|cc|}
\hline
\multicolumn{2}{|c|}{\multirow{2}{*}{}}                                                                                                                 & \multicolumn{2}{c|}{Output Difference}                                                                                                        \\ \cline{3-4} 
\multicolumn{2}{|c|}{}                                                                                                                                  & \multicolumn{1}{c|}{$0$}                                              & $1$                                                                   \\ \hline
\multicolumn{1}{|c|}{\multirow{18}{*}{\begin{tabular}[c]{@{}c@{}}I\\ n\\ p\\ u\\ t\\ \\ D\\ i\\ f\\ f\\ e\\ r\\ e\\ n\\ c\\ e\end{tabular}}} & 0        & $2^n$                                                                 & 0                                                                     \\
\multicolumn{1}{|c|}{}                                                                                                                       & 1        & 0                                                                     & $2^n$                                                                 \\
\multicolumn{1}{|c|}{}                                                                                                                       & 2        & 0                                                                     & $2^n$                                                                 \\
\multicolumn{1}{|c|}{}                                                                                                                       & 3        & $2^n$                                                                 & 0                                                                     \\
\multicolumn{1}{|c|}{}                                                                                                                       & 4        & 0                                                                     & $2^n$                                                                 \\
\multicolumn{1}{|c|}{}                                                                                                                       & 5        & $2^n$                                                                 & 0                                                                     \\
\multicolumn{1}{|c|}{}                                                                                                                       & 6        & $2^n$                                                                 & 0                                                                     \\
\multicolumn{1}{|c|}{}                                                                                                                       & 7        & 0                                                                     & $2^n$                                                                 \\
\multicolumn{1}{|c|}{}                                                                                                                       & 8        & 0                                                                     & $2^n$                                                                 \\
\multicolumn{1}{|c|}{}                                                                                                                       & 9        & $2^n$                                                                 & 0                                                                     \\
\multicolumn{1}{|c|}{}                                                                                                                       & 10       & $2^n$                                                                 & 0                                                                     \\
\multicolumn{1}{|c|}{}                                                                                                                       & 11       & 0                                                                     & $2^n$                                                                 \\
\multicolumn{1}{|c|}{}                                                                                                                       & 12       & $2^n$                                                                 & 0                                                                     \\
\multicolumn{1}{|c|}{}                                                                                                                       & 13       & 0                                                                     & $2^n$                                                                 \\
\multicolumn{1}{|c|}{}                                                                                                                       & 14       & 0                                                                     & $2^n$                                                                 \\
\multicolumn{1}{|c|}{}                                                                                                                       & 15       & $2^n$                                                                 & 0                                                                     \\
\multicolumn{1}{|c|}{}                                                                                                                       & $\vdots$ & $\vdots$                                                              & $\vdots$                                                              \\
\multicolumn{1}{|c|}{}                                                                                                                       & $2^n-1$  & $\begin{cases}2^n & n\text{ is even}\\0 & n\text{ is odd}\end{cases}$ & $\begin{cases}0 & n\text{ is even}\\2^n & n\text{ is odd}\end{cases}$ \\ \hline
\end{tabular}

\end{table}

In Equation~\ref{eq:xor-better-model} the range of values $e$ can take as,
\begin{align}
    0\leq e\leq\left\lfloor\dfrac{n+1}{2}\right\rfloor\label{eq:dummy-bounds}
\end{align}

Since the selection of values for the variables on the left-hand side of Eqution~\ref{eq:xor-better-model} completely
determines $e$, we omit to specify these constraints. Correctness of this model is trivial: it only admits solutions corresponding to non-zero cells in the DDT of an $n$-input \exor gate (see Table~\ref{tab:nxor-ddt}). We also compare the weight of this model with the previous one in Table~\ref{tab:xor-comp}. Ours always introduces only one more variable and constraint, irrespective of the size of the \exor gate.

\begin{table}[htbp]
\centering
\caption{Comparison of $n$-input \exor gate models}
\label{tab:xor-comp}
\begin{tabular}{|c|cc|cc|}
\hline
\multirow{2}{*}{\begin{tabular}[c]{@{}c@{}}Number \\ of inputs\end{tabular}} & \multicolumn{2}{c|}{\begin{tabular}[c]{@{}c@{}}Number of\\ New variables\end{tabular}} & \multicolumn{2}{c|}{\begin{tabular}[c]{@{}c@{}}Number of\\ New constraints\end{tabular}} \\ \cline{2-5} 
& \multicolumn{1}{c|}{Sasaki et al.}                         & Ours                         & \multicolumn{1}{c|}{Sasaki et al.}                          & Ours                          \\ \hline
2                                                                            & \multicolumn{1}{c|}{0}                                     & 1                            & \multicolumn{1}{c|}{4}                                      & 1                             \\ \hline
3                                                                            & \multicolumn{1}{c|}{0}                                     & 1                            & \multicolumn{1}{c|}{8}                                      & 1                             \\ \hline
4                                                                            & \multicolumn{1}{c|}{0}                                     & 1                            & \multicolumn{1}{c|}{16}                                     & 1                             \\ \hline
5                                                                            & \multicolumn{1}{c|}{0}                                     & 1                            & \multicolumn{1}{c|}{32}                                     & 1                             \\ \hline
6                                                                            & \multicolumn{1}{c|}{0}                                     & 1                            & \multicolumn{1}{c|}{64}                                     & 1                             \\ \hline
7                                                                            & \multicolumn{1}{c|}{0}                                     & 1                            & \multicolumn{1}{c|}{128}                                    & 1                             \\ \hline
\end{tabular}

\end{table}

It is worth noting that Yu Sasaki and Yosuke Todo~\cite{sasakililliput} made a passing reference to modelling large \exor gates in this manner (but with the bounds, as in Equation~ \ref{eq:dummy-bounds}) without formally laying out the concept and defining why it would work, as we have done here.

\subsection{Implementation and results}
As a stress test, we use the above two models to simulate 3200 $n$-input \exor gates constructed such that from a 16-bit input  $(u_{(0,0)}u_{(0,1)}u_{(0,2)}\ldots u_{(0,15)})$
a 16-bit output can be produced $(u_{(200,0)}u_{(200,1)}u_{(200,2)} \cdots u_{(200,15)})$
according to the following relation,
\begin{align*}
    u_{(r+1,i)}=\bigoplus_{j=i}^{i+n}u_{(r,j\bmod16)}, \forall i\in\{0,1,2,\cdots,15\}
\end{align*}
where $r\in\{0,1,2,\cdots,199\}$. We solve each model a total of 16 times (once for each input and output nibble pair
constrained to have non-zero differences).

\begin{table}[htbp]
\centering
   \caption{Performance of $n$-input \exor gate models}
    \label{tab:xor-perf}
\begin{tabular}{|c|cc|}
\hline
\multirow{2}{*}{\begin{tabular}[c]{@{}c@{}}Number of\\ Inputs\end{tabular}} & \multicolumn{2}{c|}{Solving Time (sec)}   \\ \cline{2-3} 
 & \multicolumn{1}{c|}{Sasaki et al} & Ours \\ \hline
2                                                                           & \multicolumn{1}{c|}{3.40}         & 1.45 \\
3                                                                           & \multicolumn{1}{c|}{3.18}         & 1.42 \\
4                                                                           & \multicolumn{1}{c|}{21.89}        & 1.54 \\
5                                                                           & \multicolumn{1}{c|}{49.84}        & 1.34 \\
6                                                                           & \multicolumn{1}{c|}{91.43}        & 1.73 \\
7                                                                           & \multicolumn{1}{c|}{252.15}       & 1.67 \\ \hline
\end{tabular}

\end{table}

The total time taken to solve the models is shown in Table~\ref{tab:xor-perf}. We perform this experiment on a machine
with a 32C/32T Intel Xeon Gold 6130 CPU running 64-bit CentOS Linux 7 (Core) using the Python API of Gurobi Optimizer\cite{Gurobi}.

\section{MILP-Aieded Autometic Search for Differential and Impossible Differential Propagations}\label{MILP_Cipher}

Differential cryptanalysis is one of the most fundamental cryptanalysis techniques. Identifying distinct differential trails is the first and most crucial step in the method. The concern of cryptographers has been on the automatic search methods for differentiating properties. In order to build an automatic search algorithm for differential cryptanalysis, MILP has been explicitly used. The goal of the MILP problem is to optimize an objective function while taking into account specific constraints. It belongs to a class of optimization problems derived from linear programming.

Impossible differential cryptanalysis is rooted in a miss-in-the-middle attack strategy, where an input difference is propagates forward through the cipher's rounds while an output difference is propagated backward. At an intermediate round, discrepancies arise between the transformed input and output differences. 

We introduce MILP-based solutions for valid differential trail propagation through the linear and non-linear layers. Applying these primitive components we designed a tool for searching differential and impossible differential propagation's through a cipher path.   
We created a tool that generates a MILP model for a user-specified number of rounds when the round function is given for an SPN block cipher. The model is then solved to discover a differential characteristic that minimises the amount of active \sbxs. It also searches impossible differential characteristics corresponding to impossible transitions from the input to the output.
Our tool have following distinguishable properties,
\begin{itemize}
    \item \textbf{Customizable linear layer }The user can change the linear layer components. It is possible to test a cipher by replacing a new linear layer. 
    \item \textbf{Customizable non-linear layer }One can apply an arbitrary \sbx to any cipher and check its differential propagations. 
    \item \textbf{Granularity option in searching an input/output difference }During searching of impossible differential propagations the user have various options to provide the input/output differences.
    \item \textbf{Faster operations }Due to choosing optimized primitive linear and components in every rounds the searching operations are fast.
\end{itemize}
For verification we applied the tool to Lilliput, GIFT64, SKINNY64, Klein and MIBS. The results for impossible differential propagations are tabulted in Table~\ref{tab:impossible_diff_char_count_summary}.

\begin{table}[htbp]
\centering
\caption{List of impossible differential characteristics}
\label{tab:impossible_diff_char_count_summary}

\begin{tabular}{|c|c|c|c|c|}
\hline
Cipher                   & Source & Rounds & Count & Execution Time \\ \hline
\multirow{2}{*}{Lilliput} & Our & 9      & 217   & 27~mins         \\ \cline{2-5}                    & Sasaki and Todo~\cite{autosearch-sasaki}& 9 & 217  & - \\ \hline
\multirow{3}{*}{GIFT64}   & Our & 5      & 7200  &  4~sec              \\ \cline{2-5} 
& Our & 6       & 768      & 45~sec               \\ \cline{2-5}
& Banik et al.\cite{GIFT_cipher} & 6       & -      & -               \\ \hline
\multirow{2}{*}{SKINNY64} & Our & 11     & 2700  & 21~sec               \\ \cline{2-5} 
& Beierle et al.~\cite{skinny} &      11  &  -     &   -             \\ \hline
\multirow{2}{*}{Klein}    & Our& 4      & 40    & 45~mins               \\ \cline{2-5}            & Han et al.~\cite{KLEIN_IMP_DIFF} &   4     &  -    &  -   \\ \hline
\multirow{2}{*}{MIBS} & Our & 8 & 6 & 2~hours 12~mins\\
\cline{2-5} &Bay et al.~\cite{DBLP:conf/cans/MIBS_IDC} & 8 & - & -\\ \hline

\end{tabular}
\end{table}

\subsection{Autometic search}
With the well-established MILP models for all three layers in place, it becomes feasible to combine them in various ways to accurately represent any cryptographic cipher.
We introduce Algorithm~\ref{al:gencmodel}, which outlines an automated procedure for exploring differential and impossible differential propagations. To accurately represent \sbx'es and PBoxes, it is essential to provide their specifications. This necessitates the establishment of a grammar for expressing these specifications. In the pursuit of efficient cryptanalysis, it is imperative that this grammar remains as straightforward as possible.

Additionally, a mechanism should be incorporated to enable users to specify whether they intend to discover a differential characteristic, characterized by the fewest active SBoxes, or impossible differential characteristics, denoting pairs of input and output differences that occur with a zero probability. The users can fine-tune the granularity of the search for impossible differential characteristics, with three distinct levels available.
\begin{itemize}
    \item \textbf{fuzzy} searches for impossible differential characteristics in which the input and output word differences take some non-zero values.
    
    \item \textbf{equal} searches for impossible differential characteristics in which the input and output word differences are equal.
    
    \item \textbf{targeted} performs a brute-force search for impossible differential characteristics over all
    combinations of input and output word differences.
\end{itemize}
If the granularity is not specified, it searches for a differential characteristic instead. Algorithm~\ref{al:autosearch} describes the automatic tool, which first select the attack type and then chooses the corresponding granularity. Now the tool generates the objective function and associate the MILP constraints. 

\begin{algorithm}[htbp]
    \begin{algorithmic}
        \State Inputs: File $F$ containing the specification of the round function of a cipher, number of rounds $n$.
        
        \Function{GenerateCipherModel}{$F,n$}
            \Repeat
                \State $n\gets n-1$
                \For{$\text{block}\in F$}
                    \If{$\text{block}=\texttt{SBox}$}
                        \State write inequalities modelling the non-linear layer
                    \ElsIf{$\text{block}=\texttt{XOR}$}
                        \State write inequalities modelling the linear layer
                    \ElsIf{$\text{block}=\texttt{PBox}$}
                        \State write equalities modelling the permutation layer
                    \EndIf
                \EndFor
            \Until{$n>0$}
        \EndFunction
    \end{algorithmic}
    \caption{Produce an MILP model for a user-specified number of rounds of a cipher}
    \label{al:gencmodel}
\end{algorithm}

\begin{algorithm}[htbp]
    \begin{algorithmic}
        \State $Inputs$: The attack type and the MILP constraints.
        \Function{AutoSearchModel}{$\text{attack\_type},\text{Constraints}$}
               \If{$\text{attack\_type}=\texttt{differential}$}
                 \State Set the objective function
                 \State Add the model constraints
                \State Solve the MILP model
                \ElsIf{$\text{attack\_type}=\texttt{impossible\_differential}$}
                \State Keep the objective function as blank
                    \If{$\text{block}=\texttt{fuzzy}$}
                        \State Initialize the input and output difference as any fixed non-zero value
                        \State Add the model constraints
                        \State Solve the MILP model
                    \ElsIf{$\text{block}=\texttt{equal}$}
                        \State Initialize the input and output difference as equal non-zero value
                        \State Add the model constraints
                        \State Solve the MILP model
                    \ElsIf{$\text{block}=\texttt{targeted}$}
                        \State Take all possible values for input and output differences
                        \State Add the model constraints
                        \State Solve the MILP model
                    \EndIf
                \EndIf    
        \EndFunction
    \end{algorithmic}
    \caption{Searching for a differential/impossible differential path}
    \label{al:autosearch}
\end{algorithm}

\begin{table}[htbp]
    \centering
        \caption{Differential characteristics of Lilliput}
    \label{tab:lilliput-cdt}
    \begin{tabular}{|c|c|c|c|}
        \hline
        Rounds&Active SBoxes&Input Difference&Output Difference\\
        \hline
         1& 0&\texttt{460A442EE0000000}&\texttt{E0000000AEC1A4A8}\\
         2& 1&\texttt{EEEEEEEFE0000000}&\texttt{00000000E0000000}\\
         3& 2&\texttt{8E00000000000080}&\texttt{0000800080002000}\\
         4& 3&\texttt{C080000000000C00}&\texttt{A0000000A6A8AAAA}\\
         5& 5&\texttt{3A33333830000000}&\texttt{90008000199ED099}\\
         6& 9&\texttt{D4EBBB24B9000F00}&\texttt{0000800089001F00}\\
         7&12&\texttt{0200C0000E080EAE}&\texttt{3333915E000010C3}\\
         8&15&\texttt{826EA00200000800}&\texttt{0000200200D06B00}\\
        \hline
    \end{tabular}

\end{table}

\begin{table}[htbp]
    \centering
     \caption{Impossible differential characteristics of Lilliput ($r=9$)}
    \label{tab:lilliput-idt}
    \begin{tabular}{|c|c|}
        \hline
        Input Difference&Output Difference\\
        \hline
        \texttt{0000000*00000000}&\texttt{00000000000*0000}\\
        \texttt{000000*000000000}&\texttt{000000000000000*}\\
        \texttt{000000*000000000}&\texttt{00000000000000*0}\\
        \texttt{0000002000000000}&\texttt{0000000000000200}\\
        \texttt{0000003000000000}&\texttt{0000000000000300}\\
        \texttt{0000008000000000}&\texttt{0000000000000800}\\
        \texttt{0000009000000000}&\texttt{0000000000000900}\\
        \texttt{000000E000000000}&\texttt{0000000000000E00}\\
        \texttt{000000F000000000}&\texttt{0000000000000F00}\\
        \texttt{000000*000000000}&\texttt{0000000000*00000}\\
        \texttt{00000*0000000000}&\texttt{000000000000000*}\\
        \texttt{00000*0000000000}&\texttt{00000000000000*0}\\
        \texttt{00000*0000000000}&\texttt{0000000000000*00}\\
        \texttt{00000*0000000000}&\texttt{0000000000*00000}\\
        \texttt{0000*00000000000}&\texttt{000000000000000*}\\
        \texttt{0000*00000000000}&\texttt{00000000000000*0}\\
        \texttt{0000700000000000}&\texttt{0000000000000700}\\
        \texttt{0000E00000000000}&\texttt{0000000000000E00}\\
        \texttt{0000*00000000000}&\texttt{0000000000*00000}\\
        \texttt{000*000000000000}&\texttt{000000000000000*}\\
        \texttt{0001000000000000}&\texttt{0000000000000010}\\
        \texttt{0001000000000000}&\texttt{0000000000000050}\\
        \texttt{0002000000000000}&\texttt{0000000000000020}\\
        \texttt{0003000000000000}&\texttt{0000000000000030}\\
        \texttt{0004000000000000}&\texttt{0000000000000040}\\
        \texttt{0005000000000000}&\texttt{0000000000000050}\\
        \texttt{0006000000000000}&\texttt{0000000000000060}\\
        \texttt{0007000000000000}&\texttt{0000000000000070}\\
        \texttt{0008000000000000}&\texttt{0000000000000080}\\
        \texttt{0009000000000000}&\texttt{0000000000000090}\\
        \texttt{000A000000000000}&\texttt{00000000000000A0}\\
        \texttt{000B000000000000}&\texttt{00000000000000B0}\\
        \texttt{000E000000000000}&\texttt{00000000000000E0}\\
        \texttt{000F000000000000}&\texttt{00000000000000F0}\\
        \texttt{000*000000000000}&\texttt{0000000000*00000}\\
        \hline
    \end{tabular}
   
\end{table}

\begin{table}[htbp]
    \centering
      \caption{Differential characteristics of Gift64}
    \label{tab:gift64-cdt}
    \begin{tabular}{|c|c|c|c|}
        \hline
        Rounds&Active SBoxes&Input Difference&Output Difference\\
        \hline
        1& 1&\texttt{F000000000000000}&\texttt{4000200010008000}\\
        2& 2&\texttt{0000C00000000000}&\texttt{2000100000000000}\\
        3& 3&\texttt{00000000000000B0}&\texttt{0000200010000000}\\
        4& 5&\texttt{00000000000000B0}&\texttt{0400020000000880}\\
        5& 7&\texttt{00000000000000F7}&\texttt{0008004000000010}\\
        6&10&\texttt{000000000000100B}&\texttt{1000080040020201}\\
        7&13&\texttt{0000090000000B0F}&\texttt{0024201000018008}\\
        8&16&\texttt{000000000000010C}&\texttt{0800000002000001}\\
        9&18&\texttt{000000000000070C}&\texttt{0008000100020004}\\
        \hline
    \end{tabular}
  
\end{table}

\begin{table}[htbp]
    \centering
      \caption{Impossible differential characteristics of Gift64 ($r=5$)}
    \label{tab:gift64-idt}
    \begin{tabular}{|c|c|}
        \hline
        Input Difference&Output Difference\\
        \hline
        \texttt{000000000000000*}&\texttt{0000000000-00000}\\
        \texttt{000000000000000*}&\texttt{000000-000000000}\\
        \texttt{00000000000000*0}&\texttt{00000000000000-0}\\
        \texttt{0000000000000*00}&\texttt{00000000000000-0}\\
        \texttt{0000000000000*00}&\texttt{00-0000000000000}\\
        \texttt{000000000000*000}&\texttt{000000-000000000}\\
        \texttt{000000000000*000}&\texttt{00-0000000000000}\\
        \texttt{00000000000*0000}&\texttt{000000000-000000}\\
        \texttt{00000000000*0000}&\texttt{00000-0000000000}\\
        \texttt{0000000000*00000}&\texttt{0000000000000-00}\\
        \texttt{0000000000*00000}&\texttt{000000000-000000}\\
        \texttt{000000000*000000}&\texttt{0000000000000-00}\\
        \texttt{000000000*000000}&\texttt{0-00000000000000}\\
        \texttt{00000000*0000000}&\texttt{00000-0000000000}\\
        \texttt{00000000*0000000}&\texttt{0-00000000000000}\\
        \texttt{0000000*00000000}&\texttt{00000000-0000000}\\
        \texttt{0000000*00000000}&\texttt{0000-00000000000}\\
        \texttt{000000*000000000}&\texttt{000000000000-000}\\
        \texttt{000000*000000000}&\texttt{00000000-0000000}\\
        \texttt{00000*0000000000}&\texttt{000000000000-000}\\
        \texttt{00000*0000000000}&\texttt{-000000000000000}\\
        \texttt{0000*00000000000}&\texttt{0000-00000000000}\\
        \texttt{0000*00000000000}&\texttt{-000000000000000}\\
        \texttt{000*000000000000}&\texttt{00000000000-0000}\\
        \texttt{000*000000000000}&\texttt{0000000-00000000}\\
        \texttt{00*0000000000000}&\texttt{000000000000000-}\\
        \texttt{00*0000000000000}&\texttt{00000000000-0000}\\
        \texttt{0*00000000000000}&\texttt{000000000000000-}\\
        \texttt{0*00000000000000}&\texttt{000-000000000000}\\
        \texttt{*000000000000000}&\texttt{0000000-00000000}\\
        \texttt{*000000000000000}&\texttt{000-000000000000}\\
        \hline
    \end{tabular}
  
\end{table}

\subsection{Implementation and results}
We implemented Algorithm~\ref{al:gencmodel} in Python on a machine with a 32C/32T Intel Xeon Gold 6130 CPU running 64-bit
CentOS Linux 7 (Core). To solve the MILP models generated, we used the Python API of Gurobi Optimizer\cite{Gurobi}. We
analysed Lilliput\cite{Lilliput}, Gift64\cite{GIFT_cipher}, Skinny64\cite{skinny}, Midori64\cite{midori} and
Klein\cite{klein}.

\subsection{Lilliput}

For differential characteristics of liliput up to eight rounds the counts of active \sbx'es obtained through this tool is tabulated in Table~\ref{tab:lilliput-cdt}). The results align with the results reported by Yu Sasaki and Yosuke Todo~\cite{sasakililliput}. Through a \textbf{targeted} search lasting 20 minutes, we identified 217 impossible differentials for nine rounds of Lilliput, all meticulously documented in Table~\ref{tab:lilliput-idt}. (Please note that an asterisk (*) denotes any non-zero nibble value). Remarkably, this count precisely matches the findings reported by  Sasaki and Y. Todo~\cite{autosearch-sasaki}. It is worth noting, however, that their search was not exhaustive and took approximately an hour to complete.

\subsection{Gift64}

\begin{table}[htbp]
    \centering
       \caption{Differential characteristics of Skinny64}
    \label{tab:skinny64-cdt}
    \begin{tabular}{|c|c|c|c|}
        \hline
        Rounds&Active SBoxes&Input Difference&Output Difference\\
        \hline
        1& 1&\texttt{000000000000000C}&\texttt{0002000000020002}\\
        2& 2&\texttt{0000000000A00000}&\texttt{0004000400000004}\\
        3& 5&\texttt{1000000700100000}&\texttt{0020000000200020}\\
        4& 8&\texttt{9000000E00700000}&\texttt{0010010000100011}\\
        5&12&\texttt{0001000000003000}&\texttt{0414041000044414}\\
        6&16&\texttt{00260400D0000000}&\texttt{0222022000022222}\\
        7&27&\texttt{060204004800B005}&\texttt{C0060440C046C402}\\
        \hline
    \end{tabular}
 
\end{table}

\begin{table}[htbp]
    \centering
       \caption{Impossible differential characteristics of Skinny64 ($r = 11$)}
    \label{tab:skinny64-idt}
    \begin{tabular}{|c|c|}
        \hline
        Input Difference&Output Difference\\
        \hline
        \texttt{000*000000000000}&\texttt{0000000-00000000}\\
        \texttt{000*000000000000}&\texttt{000000-000000000}\\
        \texttt{000*000000000000}&\texttt{0000-00000000000}\\
        \texttt{00*0000000000000}&\texttt{0000000-00000000}\\
        \texttt{00*0000000000000}&\texttt{000000-000000000}\\
        \texttt{00*0000000000000}&\texttt{00000-0000000000}\\
        \texttt{0*00000000000000}&\texttt{000000-000000000}\\
        \texttt{0*00000000000000}&\texttt{00000-0000000000}\\
        \texttt{0*00000000000000}&\texttt{0000-00000000000}\\
        \texttt{*000000000000000}&\texttt{0000000-00000000}\\
        \texttt{*000000000000000}&\texttt{00000-0000000000}\\
        \texttt{*000000000000000}&\texttt{0000-00000000000}\\
        \hline
    \end{tabular}
 
\end{table}

Our tool successfully determined the number of active \sbx'es in Gift64 upto nine rounds, and the outcomes are consistent with the findings reported by the Gift64~\cite{GIFT_cipher} authors. Additionally, as a demonstration of the tool's capabilities, we conducted \textbf{fuzzy} impossible differential cryptanalysis on a five-round Gift64 variant. The results, encompassing 7200 characteristics, are meticulously presented in Table~\ref{tab:gift64-idt}. (In this representation, an asterisk (*) signifies any non-zero nibble value, while a hyphen (-) denotes any independently chosen non-zero nibble value.)

\subsection{Skinny64}

\begin{table}[htbp]
    \centering
     \caption{Differential characteristics of Midori64}
    \label{tab:midori64-cdt}
    \begin{tabular}{|c|c|c|c|}
        \hline
        Rounds&Active SBoxes&Input Difference&Output Difference\\
        \hline
        1& 1&\texttt{000000000000000D}&\texttt{000000000000AAA0}\\
        2& 4&\texttt{0100000000000010}&\texttt{4440000044040000}\\
        3& 7&\texttt{0000000140000100}&\texttt{0000101188800888}\\
        4&16&\texttt{0101011001110011}&\texttt{1011110100001110}\\
        5&23&\texttt{000001000010000E}&\texttt{1011444022020302}\\
        6&32&\texttt{0106000B00142004}&\texttt{4404202244400000}\\
        7&40&\texttt{0000000000080002}&\texttt{000051418C4C0000}\\
        \hline
    \end{tabular}
   
\end{table}

\begin{table}[htbp]
    \centering
       \caption{Differential characteristics of Klein}
    \label{tab:klein-cdt}
    \begin{tabular}{|c|c|c|c|}
        \hline
        Rounds&Active SBoxes&Input Difference&Output Difference\\
        \hline
        1& 1&\texttt{0000000000000009}&\texttt{0DDED30D00000000}\\
        2& 5&\texttt{0000000000000808}&\texttt{D50D07D202060404}\\
        3& 8&\texttt{0505000000000005}&\texttt{0505DADF03DEDD05}\\
        4&15&\texttt{020D000000000000}&\texttt{01D8D90103020100}\\
        5&17&\texttt{00A0000000000000}&\texttt{C0B8783000000000}\\
        \hline
    \end{tabular}
 
\end{table}

Table~\ref{tab:skinny64-cdt} presents the count of active \sbx'es for Skinny64~\cite{skinny} up to seven rounds as determined by our tool. It's worth noting that the authors reported 26 active \sbx'es for a seven-round Skinny64, whereas our tool identified 27 active \sbx'es. The difference can likely be attributed to the authors' utilization of N. Mouha's \sbx modeling~\cite{Mouha}, as opposed to employing the more stringent models.

Furthermore, we showcase a collection of 2700 impossible differential characteristics obtained through a \texttt{fuzzy} search in Table~\ref{tab:skinny64-idt}. (In this representation, an asterisk (*) signifies any non-zero nibble value, while a hyphen (-) denotes any independently chosen non-zero nibble value.)

\begin{table}[htbp]
    \centering
       \caption{Impossible differential characteristics of Klein ($r=4$)}
    \label{tab:klein-idt}
    \scalebox{0.9}{
    \begin{tabular}{|c|c|}
        \hline
        Input Difference&Output Difference\\
        \hline
        \texttt{0000000000000800}&\texttt{0000000080000000}\\
        \texttt{0000000000000B00}&\texttt{0000000B00000000}\\
        \texttt{0000000000000800}&\texttt{8000000000000000}\\
        \texttt{0000000000003000}&\texttt{0000030000000000}\\
        \texttt{0000000000007000}&\texttt{0000000000000007}\\
        \texttt{0000000000008000}&\texttt{0000000000000008}\\
        \texttt{0000000000008000}&\texttt{0000080000000000}\\
        \texttt{0000000000009000}&\texttt{0009000000000000}\\
        \texttt{000000000000B000}&\texttt{000B000000000000}\\
        \texttt{000000000000E000}&\texttt{000E000000000000}\\
        \texttt{0000000008000000}&\texttt{0000000000008000}\\
        \texttt{0000000008000000}&\texttt{0000800000000000}\\
        \texttt{000000000B000000}&\texttt{000B000000000000}\\
        \texttt{0000000030000000}&\texttt{0300000000000000}\\
        \texttt{0000000070000000}&\texttt{0000000000070000}\\
        \texttt{0000000080000000}&\texttt{0000000000080000}\\
        \texttt{0000000080000000}&\texttt{0800000000000000}\\
        \texttt{0000000090000000}&\texttt{0000000900000000}\\
        \texttt{00000000B0000000}&\texttt{0000000B00000000}\\
        \texttt{00000000E0000000}&\texttt{0000000E00000000}\\
        \texttt{0000080000000000}&\texttt{0000000080000000}\\
        \texttt{0000080000000000}&\texttt{8000000000000000}\\
        \texttt{00000B0000000000}&\texttt{000000000000000B}\\
        \texttt{0000300000000000}&\texttt{0000000000000300}\\
        \texttt{0000700000000000}&\texttt{0000000700000000}\\
        \texttt{0000800000000000}&\texttt{0000000000000800}\\
        \texttt{0000800000000000}&\texttt{0000000800000000}\\
        \texttt{0000900000000000}&\texttt{0000000000090000}\\
        \texttt{0000B00000000000}&\texttt{00000000000B0000}\\
        \texttt{0000E00000000000}&\texttt{00000000000E0000}\\
        \texttt{0800000000000000}&\texttt{0000000000008000}\\
        \texttt{0800000000000000}&\texttt{0000800000000000}\\
        \texttt{0B00000000000000}&\texttt{00000000000B0000}\\
        \texttt{3000000000000000}&\texttt{0000000003000000}\\
        \texttt{7000000000000000}&\texttt{0007000000000000}\\
        \texttt{8000000000000000}&\texttt{0000000008000000}\\
        \texttt{8000000000000000}&\texttt{0008000000000000}\\
        \texttt{9000000000000000}&\texttt{0000000000000009}\\
        \texttt{B000000000000000}&\texttt{000000000000000B}\\
        \texttt{E000000000000000}&\texttt{000000000000000E}\\
        \hline
    \end{tabular}}
 
\end{table}

\subsection{Midori64}
Regarding Midori64~\cite{midori}, we have documented the minimum counts of active \sbx'es in Table~\ref{tab:midori64-cdt}. Specifically, for six and seven rounds of Midori64, our analysis yielded 32 and 40 active \sbx'es, respectively. In contrast, the authors Midori64~\cite{midori} reported counts of 30 and 35 active \sbx'es for the same rounds. It's important to note that the authors derived their counts not through the MILP method but by pursuing optimal permutations of the rows in which the cipher state is organized. Our counts for fewer rounds align with those of the authors. Furthermore, following a fuzzy search, our tool can not identify any impossible differentials for eight rounds or more.

\subsection{Klein}
Due to the intricate nature of Klein~\cite{klein}, stemming partly from the linear layer that involves matrix multiplication with elements interpreted as polynomials in $\mathbb{F}_2^8$ (modeled according to the approach outlined by B. Ilter and A. A. Selçuk~ \cite{xormodel-chull-kleinprince}), we were able to ascertain the count of minimally active \sbx'es in Klein only for a limited number of rounds, as reflected in Table~\ref{tab:klein-cdt}. Similarly, owing to the same complexities, we have provided \textbf{equal} impossible differentials for four rounds of Klein in Table~\ref{tab:klein-idt}. The total count of impossible differential characteristics discovered in this context is 40.

\section{Conclusion}\label{Conclusion}
In this paper, we present two innovative techniques for identifying the minimum set of inequalities to model the differential propagations of an SBox. The algorithms we propose for modeling the Difference Distribution Table (DDT) of an SBox demonstrate superior efficiency compared to existing methods. Recognizing that a greedy algorithm is only complete with the incorporation of a tiebreaker, we have developed a new version of the greedy approach employing a random tiebreaker. The results of the greedy random tiebreaker surpass those of the original greedy method for certain SBoxes. The subset addition algorithm proves effective in modeling SBoxes up to 6 bits, providing better or nearly identical outcomes for most SBoxes. Additionally, we have enhanced the execution time required to identify minimized inequalities compared to previous implementations.
We have explored a concise representation of the linear layer, primarily utilizing \exor gates. This approach harnesses the capacity of MILP solvers to handle not only binary variables but also non-binary integral variables. Notably, this model not only achieves a streamlined form but also outperforms alternative models in terms of computational efficiency.

Furthermore, we have developed, implemented, and demonstrated the functionality of a tool that can interpret this specification and autonomously embark on the quest for a differential characteristic (i.e., the minimum count of differentially active \sbx'es) or impossible differential characteristics (i.e., pairs of input and output differences with a probability of zero) for any given number of rounds within a block cipher. 

%
%
%
\bibliographystyle{splncs04}
\bibliography{main}
%


\end{document}